\documentclass[prb,superscriptaddress,reprint]{revtex4-1}
\usepackage{graphicx}
\usepackage{bm}
\usepackage{ulem}
\usepackage{color}
\usepackage{amsmath}
\usepackage{amssymb}

\makeatletter
\renewcommand{\fnum@figure}{\textbf{Figure \thefigure}}
\makeatother

\begin{document}

\title{Resolving magnon number states in quantum magnonics}

\author{Dany Lachance-Quirion}
\email{dany.lachance.quirion@usherbrooke.ca}
\affiliation{Institut quantique and D\'epartement de Physique, Universit\'e de Sherbrooke, J1K 2R1, Sherbrooke, Qu\'ebec,  Canada}
\affiliation{Research Center for Advanced Science and Technology (RCAST), The University of Tokyo, Meguro-ku, Tokyo 153-8904, Japan}
\author{Yutaka Tabuchi}
\author{Seiichiro Ishino}
\author{Atsushi Noguchi}
\author{Toyofumi Ishikawa}
\author{Rekishu Yamazaki}
\affiliation{Research Center for Advanced Science and Technology (RCAST), The University of Tokyo, Meguro-ku, Tokyo 153-8904, Japan}
\author{Yasunobu Nakamura}
\email{yasunobu@qc.rcast.u-tokyo.ac.jp}
\affiliation{Research Center for Advanced Science and Technology (RCAST), The University of Tokyo, Meguro-ku, Tokyo 153-8904, Japan}
\affiliation{Center for Emergent Matter Science (CEMS), RIKEN, Wako, Saitama 351-0198, Japan}

\date{October 4$^\mathrm{th}$ 2016}

\maketitle

\textbf{Collective excitation modes in solid state systems play a central role in circuit quantum electrodynamics~\cite{Blais2004,Wallraff2004a}, cavity optomechanics~\cite{Aspelmeyer2014}, and quantum magnonics~\cite{Tabuchi2015a,Tabuchi2016}. In the latter, quanta of collective excitation modes in a ferromagnet, called magnons, interact with qubits to provide the nonlinearity necessary to access quantum phenomena in magnonics. A key ingredient for future quantum magnonics systems is the ability to probe magnon states. Here we observe individual magnons in a millimeter-sized ferromagnet coherently coupled to a superconducting qubit. Specifically, we resolve magnon number states in spectroscopic measurements of a transmon qubit with the hybrid system in the strong dispersive regime~\cite{Gambetta2006,Schuster2007}. This enables us to detect a change in the magnetic dipole of the ferromagnet equivalent to a single spin flipped among more than $10^{19}$ spins. The strong dispersive regime of quantum magnonics opens up the possibility of encoding superconducting qubits into non-classical magnon states~\cite{Leghtas2013a,Vlastakis2013a}, potentially providing a coherent interface between a superconducting quantum processor and optical photons~\cite{Kimble2008,Osada2016,Hisatomi2016,Zhang2016,Haigh2016}.}

Cavity and circuit quantum electrodynamics have enabled the realization of many gedanken experiments in quantum optics~\cite{Haroche2006,Hofheinz2009,Kirchmair2013}, as well as offering a promising platform for quantum computing~\cite{Ladd2010,Kelly2015}. Ideas from these fields have been applied to cavity optomechanics~\cite{Aspelmeyer2014,Teufel2011,Lecocq2015}, with phonons in mechanical resonators, and quantum magnonics~\cite{Tabuchi2016}, with magnons in magnetostatic modes. The coherent interaction between magnons and a superconducting qubit was demonstrated through the observation of a magnon-vaccum Rabi splitting of the qubit~\cite{Tabuchi2015a}. While this experiment was performed when both systems are resonantly coupled, here we go further and explore the off-resonant, dispersive regime of quantum magnonics, a promising regime to probe magnon states with the qubit.

\begin{figure*}[htb!]
\includegraphics[scale=0.95]{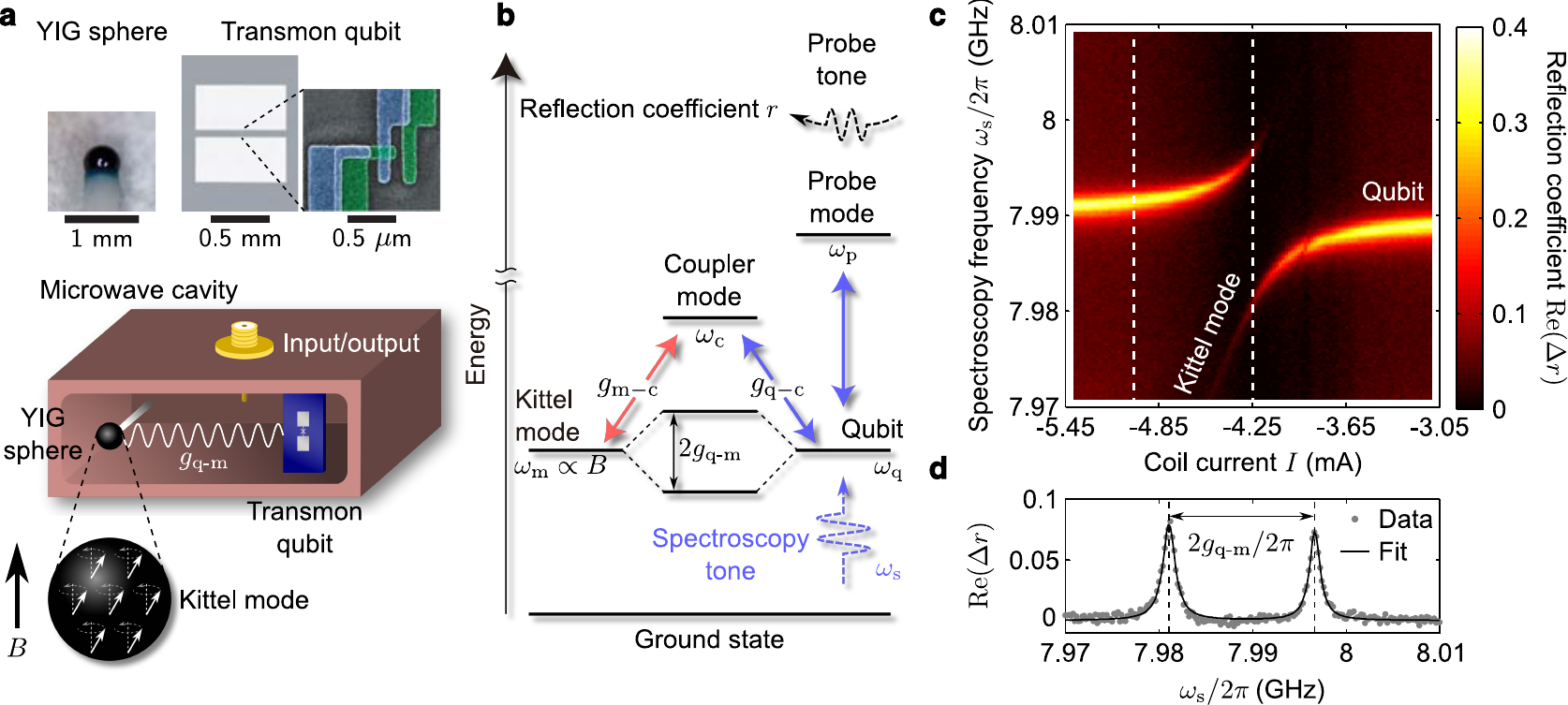}
\caption{\textbf{Hybrid system and qubit-magnon coherent interaction.} \textbf{a},~Schematic illustration of a yttrium iron garnet (YIG) sphere and a superconducting transmon qubit inside a three-dimensional microwave cavity. A magnetic field $B$ is applied to the YIG sphere using permanent magnets and a coil. The magnetostatic mode in which spins precess uniformly in the ferromagnetic sphere, or the Kittel mode, couples to the magnetic field of the cavity modes. The Kittel mode and the transmon qubit interact through virtual excitations in the cavity modes at a rate $g_\mathrm{q-m}$. \textbf{b},~Schematic energy diagram of the coupler and probe cavity modes, the qubit, and the Kittel mode, respectively of frequencies $\omega_\mathrm{c}$, $\omega_\mathrm{p}$, $\omega_\mathrm{q}$, and $\omega_\mathrm{m}$. The qubit and the Kittel mode are coupled to the coupler cavity mode through electric and magnetic dipole interactions, at rates $g_\mathrm{q-c}$ and $g_\mathrm{m-c}$, respectively. The spectrum of the qubit is measured by probing reflection of a microwave excitation, resonant with the probe mode at frequency $\omega_\mathrm{p}$, as a function of the spectroscopy frequency $\omega_\mathrm{s}$. \textbf{c},~Measurement of the qubit spectrum by probing the change of the reflection coefficient, $\mathrm{Re}(\Delta r)$, as a function of $\omega_\mathrm{s}$ and coil current $I$, changing the magnetic field at the YIG sphere. The avoided crossing indicates a coherent interaction between the qubit and the Kittel mode. Vertical dashed lines indicate $I=-4.25$~mA, where the qubit and the Kittel mode are hybridized~(Fig.~\ref{fig:Figure_1}d), and $I=-5.02$~mA, where the qubit-magnon interaction is in the dispersive regime~(Figs.~\ref{fig:Figure_2} and \ref{fig:Figure_3}). \textbf{d},~Magnon-vacuum Rabi splitting of the qubit on resonance with the Kittel mode at $I=-4.25$~mA. From the fit, we extract the qubit-magnon coupling rate $g_\mathrm{q-m}/2\pi=7.79$~MHz.
\label{fig:Figure_1}}
\end{figure*}

To this end, we use an hybrid system which consists of a superconducting qubit and a single-crystalline yttrium iron garnet (YIG) sphere inside a three-dimensional microwave cavity~(Fig.~\ref{fig:Figure_1}a). The transmon-type superconducting qubit has a resonant frequency of 7.9905~GHz. A pair of permanent magnets and a coil are used to apply a magnetic field $B$ to the YIG sphere, making it a single-domain ferromagnet. The electric dipole of the qubit and the magnetic dipole of the ferromagnet couple to the electric and magnetic fields of the cavity modes, respectively. The YIG sphere is placed near the anti-node of the magnetic field of the transverse electric TE$_{102}$ cavity mode, so that the cavity field is nearly uniform throughout the 0.5-mm sphere. This makes the uniformly-precessing mode, or Kittel mode, the most dominantly coupled magnetostatic mode. The interaction strength $g_\mathrm{m-c}$ between the TE$_{102}$ cavity mode, or coupler mode, and the Kittel mode reaches the strong coupling regime~\cite{Huebl2013,Tabuchi2014,Goryachev2014a,Zhang2015c,Lambert2016}, where $g_\mathrm{m-c}/2\pi=22.5$~MHz is much larger than both the cavity and magnon linewidths, respectively $\kappa_\mathrm{c}/2\pi=2.08$~MHz and $\gamma_\mathrm{m}/2\pi=1.3$~MHz~(Supplementary Information).

The coupling of the superconducting qubit and the Kittel mode to the same cavity modes creates an effective interaction between these two macroscopic systems~\cite{Tabuchi2015a,Tabuchi2016}. To verify that this interaction is coherent, we perform spectroscopic measurements of the qubit with the hybrid system in the quantum regime at a temperature of 10~mK. While the qubit-magnon coupling is mostly provided by the TE$_{102}$ cavity mode at 8.4563~GHz (coupler mode), we use the dispersive interaction of the qubit with the TE$_{103}$ cavity mode at 10.4492~GHz (probe mode) for reading out the qubit~(Fig.~\ref{fig:Figure_1}b). This scheme avoids measurement-induced dephasing caused by photon number fluctuations in the coupler mode~\cite{Gambetta2006,Tabuchi2016}. The change in the reflection coefficient $r$ of a probe microwave tone, resonant with the probe mode and containing less than one photon on average, is measured while exciting the qubit with a spectroscopic microwave tone at frequency $\omega_\mathrm{s}$. Figure~\ref{fig:Figure_1}c shows the real part of the change of the reflection coefficient, $\Delta r$, measured for different currents $I$ in the coil, changing the frequency $\omega_\mathrm{m}\propto B$ of the magnons in the Kittel mode. The avoided crossing indicates the coherent interaction between the qubit and the Kittel mode~\cite{Tabuchi2015a}. Indeed, the qubit-magnon coupling strength $g_\mathrm{q-m}$ of 7.79~MHz, obtained from the magnon-vacuum Rabi splitting of the qubit~(Fig.~\ref{fig:Figure_1}d), is much larger than both the power-broadened qubit linewidth $\gamma_\mathrm{q}/2\pi=1.74$~MHz and the magnon linewidth $\gamma_\mathrm{m}/2\pi=1.3$~MHz.

\begin{figure}[htb!]
\includegraphics[scale=0.95]{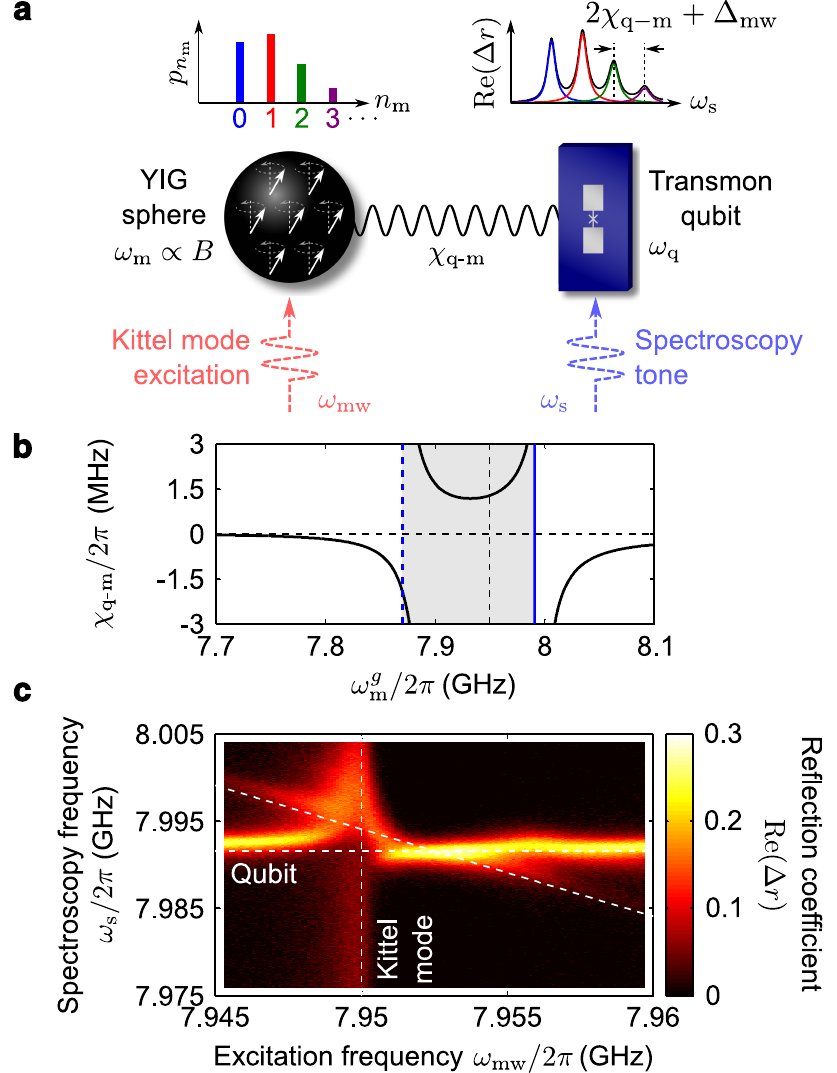}
\caption{\textbf{Dispersive qubit-magnon interaction.} \textbf{a},~Schematic illustration of the hybrid system in the strong dispersive regime. A microwave excitation at frequency $\omega_\mathrm{mw}$ is used to create a magnon coherent state in the Kittel mode. The excitation is detuned from the magnon frequency with the qubit in the ground state, $\omega_\mathrm{m}^g$, by $\Delta_\mathrm{mw}=\omega_\mathrm{m}^g-\omega_\mathrm{mw}$. In the strong dispersive regime, magnon number states $|n_\mathrm{m}\rangle$ (of probability distribution $p_{n_\mathrm{m}}$) are mapped into the qubit spectrum as peaks at frequencies $\omega_\mathrm{q}^{(n_\mathrm{m})}$, separated by $2\chi_\mathrm{q-m}+\Delta_\mathrm{mw}$ and with a spectral weight closely related to $p_{n_\mathrm{m}}$. \textbf{b},~Calculation of the qubit-magnon dispersive shift $\chi_\mathrm{q-m}$ as a function of the magnon frequency with the qubit in the ground state, $\omega_\mathrm{m}^g$. The straddling regime (shaded area) corresponds to $\omega_\mathrm{m}^g$ between the transmon $|g\rangle\leftrightarrow|e\rangle$ transition frequency ($\omega_\mathrm{q}$, blue solid line) and $|e\rangle\leftrightarrow|f\rangle$ transition frequency ($\omega_\mathrm{q}+\alpha$, blue dashed line, where $\alpha(<0)$ is the transmon anharmonicity) with the Kittel mode in the vacuum state. \textbf{c},~Measurement of the qubit spectrum for a coil current $I=-5.02$~mA as a function of the Kittel mode excitation frequency $\omega_\mathrm{mw}$ and the spectroscopy frequency $\omega_\mathrm{s}$. The excitation frequency producing the maximum magnon-induced ac Stark shift of the qubit from $\omega_\mathrm{q}$ (horizontal dashed line) yields an estimation of $\omega_\mathrm{m}^g/2\pi\approx7.95$~GHz (vertical dashed line) for $I=-5.02$~mA. The Kittel mode spectrum, measured via its dispersive interaction with the probe mode, appears as a faint vertical line at $\sim7.95$~GHz. Finally, the qubit transition frequency with a single magnon in the Kittel mode is shown as a diagonal dashed line calculated with the expected value $\chi_\mathrm{q-m}/2\pi=1.27$~MHz at $\omega_\mathrm{m}^g/2\pi=7.95$~GHz (vertical dashed line in \textbf{b}). 
\label{fig:Figure_2}}
\end{figure}

We now investigate the dispersive regime of our hybrid system, where the detuning between the bare qubit and Kittel mode frequencies, $\left|\omega_\mathrm{q}^\mathrm{bare}-\omega_\mathrm{m}^\mathrm{bare}\right|$, is much larger than $g_\mathrm{q-m}$. The exchange of quanta of excitations between the qubit and the Kittel mode, through virtual photons in the coupler cavity mode, is then highly suppressed. The dispersive part of the qubit-magnon Hamiltonian is then given by
\begin{align}
\mathcal{\hat H}_\mathrm{q-m}^\mathrm{disp}=\hbar\chi_\mathrm{q-m}\hat \sigma_z\hat c^\dagger \hat c,
\label{eq:Hamiltonian}
\end{align}
where $\hat\sigma_z=|e\rangle\langle e|-|g\rangle\langle g|$, with $|g(e)\rangle$ the ground (excited) state of the transmon qubit, $\hat c^\dagger$ ($\hat c$) is the magnon creation (annihilation) operator, and $\chi_\mathrm{q-m}$ is the qubit-magnon dispersive shift~\cite{Blais2004,Tabuchi2016}. This dispersive interaction makes the qubit and magnon frequencies dependent on the state of the other system. More precisely, the qubit frequency $\omega_\mathrm{q}^{(n_\mathrm{m})}$ depends on the magnon number state $|n_\mathrm{m}=\{0,1,2,\dots\}\rangle$ and the magnon frequency $\omega_\mathrm{m}^i$ depends on the transmon state $|i=\{g,e,f,\dots\}\rangle$. As illustrated in Fig.~\ref{fig:Figure_2}a, the strong dispersive regime, where $\left|2\chi_\mathrm{q-m}\right|>\mathrm{max}\left[\gamma_\mathrm{q},\gamma_\mathrm{m}\right]$, enables the observation of magnon number states $|n_\mathrm{m}\rangle$ via magnon-number-dependent ac Stark shift of the qubit frequency when the Kittel mode is coherently driven~\cite{Gambetta2006,Schuster2007}. Figure~\ref{fig:Figure_2}b shows the dispersive shift $\chi_\mathrm{q-m}$ expected for our hybrid system as a function of the dressed magnon frequency $\omega_\mathrm{m}^g$. The straddling regime, where we perform the spectroscopic study, corresponds to $\omega_\mathrm{q}+\alpha<\omega_\mathrm{m}^g<\omega_\mathrm{q}$, where $\omega_\mathrm{q}$ is the qubit frequency with the Kittel mode in the vacuum state. The transmon anharmonicity, $\alpha/2\pi=-120.2$~MHz, is defined with $\omega_\mathrm{q}+\alpha$ as the transition frequency between the $|e\rangle$ and $|f\rangle$ states of the transmon~\cite{Koch2007,Inomata2012a} (Supplementary Information).

As illustrated in Fig.~\ref{fig:Figure_2}a, the qubit-magnon dispersive regime is investigated through spectroscopic measurements of the qubit while exciting the Kittel mode at frequency $\omega_\mathrm{mw}$, detuned by $\Delta_\mathrm{mw}=\omega_\mathrm{m}^g-\omega_\mathrm{mw}$ from the dressed magnon frequency $\omega_\mathrm{m}^g$. The measurement of the qubit spectrum while sweeping $\omega_\mathrm{mw}$ for a coil current of $-5.02$~mA and a Kittel mode excitation power $P_\mathrm{mw}$ of $7.9$~fW is shown in Fig.~\ref{fig:Figure_2}c. Near resonant excitation $\Delta_\mathrm{mw}\sim0$, the qubit is ac Stark shifted by the magnon occupancy in the Kittel mode, a signature of the qubit-magnon dispersive interaction similar to the qubit-photon counterpart in circuit QED experiments~\cite{Schuster2005,Gambetta2006}. The positive magnon-induced ac Stark shift shows that $\chi_\mathrm{q-m}>0$, while the excitation frequency producing the maximum shift indicates $\omega_\mathrm{m}^g/2\pi\approx7.95$~GHz. Both these features are consistent with the hybrid system being in the straddling regime for $I=-5.02$~mA~(Fig.~\ref{fig:Figure_2}b). More importantly, the peak corresponding to the qubit $|g\rangle\leftrightarrow|e\rangle$ transition with a \textit{single} magnon in the Kittel mode, shifted from the zero-magnon qubit transition by $2\chi_\mathrm{q-m}+\Delta_\mathrm{mw}$, is visible at large detunings.
	
\begin{figure*}[htb!]
\includegraphics[scale=0.95]{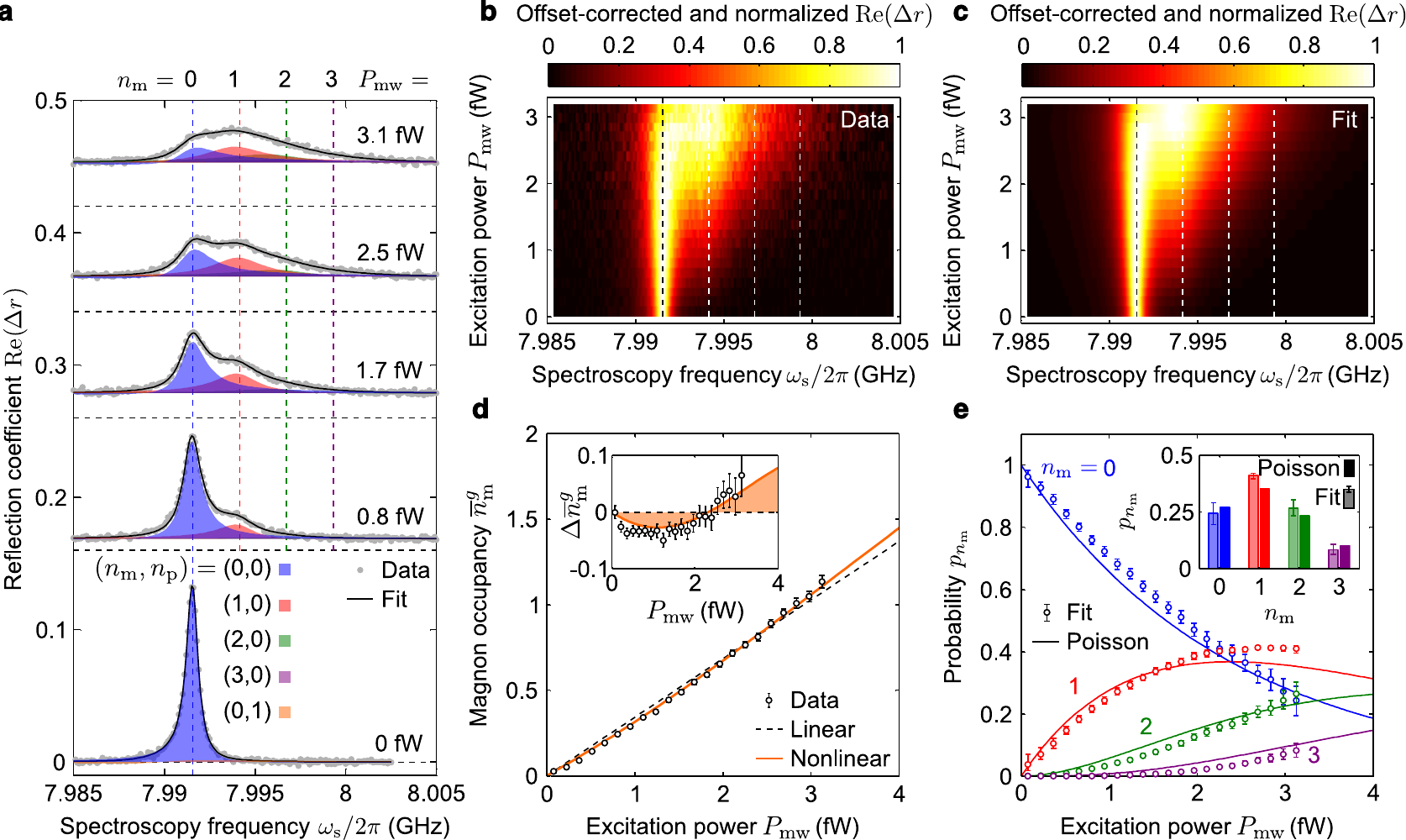}
\caption{\textbf{Resolving magnon number states.} \textbf{a},~Measurements of the qubit spectrum at different Kittel mode excitation powers $P_\mathrm{mw}$ for a coil current of $-5.02$~mA and Kittel mode excitation frequency of 7.95~GHz. Black lines show fits of the data to the spectrum of a qubit dispersively coupled to a harmonic oscillator. Color-coded shaded areas show components of the spectrum corresponding to different photon number states $|n_\mathrm{p}\rangle$ of the probe cavity mode and magnon number states $|n_\mathrm{m}\rangle$. Vertical offsets are shown by horizontal dashed lines. \textbf{b},~Measured and \textbf{c},~fitted qubit spectra as a function of $P_\mathrm{mw}$. For clarity, after subtracting a power-dependent offset, $\mathrm{Re}(\Delta r)$ is normalized relative to its maximum for each drive power. For \textbf{a}, \textbf{b} and \textbf{c}, vertical dashed lines indicate the frequencies $\omega_\mathrm{q}^{(n_\mathrm{m})}$ of the dressed qubit $|g\rangle\leftrightarrow|e\rangle$ transitions corresponding to the first four magnon number states. \textbf{d},~Magnon occupancy $\overline{n}_\mathrm{m}^g$ as a function of the excitation power. Dashed black line and solid orange line show fits to a driven linear and nonlinear Kittel mode, respectively. Inset: Deviations $\Delta\overline{n}_\mathrm{m}^g$ from the linear fit, indicating a significant magnon Kerr nonlinearity. \textbf{e},~Probability~$p_{n_\mathrm{m}}$ of the first four magnon number states as a function of $P_\mathrm{mw}$. Poisson distributions are shown as solid lines. Inset: $p_{n_\mathrm{m}}$ for $P_\mathrm{mw}=3.1$~fW. For both \textbf{d} and \textbf{e}, error bars correspond to 95\% confidence intervals.
\label{fig:Figure_3}}
\end{figure*}

We now focus on resolving magnon number states through measurements with the excitation frequency close to resonance with the Kittel mode ($\Delta_\mathrm{mw}\ll\gamma_\mathrm{m}$). In the qubit spectra shown in Figs.~\ref{fig:Figure_3}a and \ref{fig:Figure_3}b, the excitation frequency is fixed at $7.95$~GHz, close to resonance with $\omega_\mathrm{m}^g$ for $I=-5.02$~mA~(Fig.~\ref{fig:Figure_2}c). The microwave excitation creates a magnon coherent state in the Kittel mode. Indeed, when coherently driving the Kittel mode, we observe peaks in the qubit spectrum at frequencies higher than the zero-magnon peak. These peaks correspond to different number of magnons in the Kittel mode.

To fit the data, we use an analytical model of the spectrum of a qubit dispersively coupled to a harmonic oscillator~\cite{Gambetta2006}. The asymmetric qubit lineshape at $P_\mathrm{mw}=0$ is well reproduced by including in the fit the photonic contribution to the qubit lineshape from the dispersive interaction between the qubit and the probe mode (Supplementary Information). The fitting parameters for each excitation power are the occupancy of the Kittel mode $\overline{n}_\mathrm{m}^g=\langle\hat n_\mathrm{m}\hat\Pi_\mathrm{q}^g\rangle$, where $\hat\Pi_\mathrm{q}^g=|g\rangle\langle g|$ is the projector to the qubit ground state, the qubit-magnon dispersive shift, $\chi_\mathrm{q-m}$, and the excitation detuning, $\Delta_\mathrm{mw}$~(Figs.~\ref{fig:Figure_3}a and \ref{fig:Figure_3}c). We find a detuning $\Delta_\mathrm{mw}$ of $-0.38$~MHz, indicating a bare magnon frequency $\omega_\mathrm{m}^\mathrm{bare}$ of $7.9515$~GHz (Supplementary Information). The condition for the dispersive regime is therefore clearly respected with a detuning $|\omega_\mathrm{q}^\mathrm{bare}-\omega_\mathrm{m}^\mathrm{bare}|$ of $89$~MHz much larger than the qubit-magnon coupling strength. The qubit-magnon dispersive shift $\chi_\mathrm{q-m}$ is found to be $1.5\pm0.1$~MHz, in good agreement with the theoretical value of $1.27$~MHz~(Fig.~\ref{fig:Figure_2}b). Resolving magnon number states clearly demonstrates that we have reached the strong dispersive regime of quantum magnonics, with the dispersive shift per magnon $2\chi_\mathrm{q-m}$ larger than both the power-broadened qubit linewidth $\gamma_\mathrm{q}/2\pi=0.78$~MHz and the magnon linewidth $\gamma_\mathrm{m}/2\pi=1.3$~MHz.

The average number of magnons $\overline{n}_\mathrm{m}^g$ in the Kittel mode extracted from the fit of the data is shown in Fig.~\ref{fig:Figure_3}d. At the lowest excitation power of 79~aW, we are able to resolve $0.026\pm0.012$ magnons in the Kittel mode of the ferromagnetic sphere containing about $1.4\times10^{18}$ spins. This is therefore a change in magnetic dipole equivalent to a single spin flipped among $\sim5\times10^{19}$ spins. Within our resolution of about 0.01 magnons, no thermal occupancy of the Kittel mode is observed, indicating that the millimeter-sized ferromagnet is indeed in its magnon-vacuum state. Furthermore, the nonlinearity of $\overline{n}_\mathrm{m}^g$ against the excitation power can be explained by a Kerr nonlinearity in the Kittel mode caused by the anharmonicity of the transmon~\cite{Bourassa2012,Kirchmair2013}. We find a Kerr coefficient of $-0.20\substack{+0.09\\ -0.13}$~MHz, in good agreement with the expected value of $-0.12$~MHz (Supplementary Information). 

Finally, we compute the probability $p_{n_\mathrm{m}}$ of having $n_\mathrm{m}$ magnons in the Kittel mode with
\begin{align}
p_{n_\mathrm{m}}=\int\mathrm{d}\omega_\mathrm{s}\ S_{n_\mathrm{m}}(\omega_\mathrm{s})\left/S(\omega_\mathrm{s})\right.,
\label{eq:prob}
\end{align}
where $S(\omega_\mathrm{s})\approx\sum_{n_\mathrm{m}=0}^{10}S_{n_\mathrm{m}}(\omega_\mathrm{s})$ is the qubit spectrum in the analytical model, to which data is fitted, and $S_{n_\mathrm{m}}(\omega_\mathrm{s})$ is its component associated with the magnon number state $|n_\mathrm{m}\rangle$. For $2\chi_\mathrm{q-m}\gg\gamma_\mathrm{m}$, the probability distribution calculated with equation~\eqref{eq:prob} falls back to the Poisson distribution expected for a driven harmonic oscillator~\cite{Gambetta2006} (Supplementary Information). The probability distributions $p_{n_\mathrm{m}}$ of the first four magnon number states, shown in Fig.~\ref{fig:Figure_3}e, indicate small deviations from Poisson distributions. This is expected as the qubit-magnon dispersive shift is only slightly larger than the magnon linewidth in our hybrid system. Nevertheless, our ability to map the probability distribution of magnon number states to the spectrum of a qubit provides a novel tool to investigate quantum states in magnetostatic modes.

Looking forward, the strong dispersive interaction between magnons and a superconducting qubit demonstrated here should enable the encoding of the qubit into a superposition of magnon coherent states in a magnetostatic mode~\cite{Leghtas2013a,Vlastakis2013a}. However, to implement this encoding protocol, the qubit-magnon system needs to be deeper into the strong dispersive regime, either by increasing the qubit-magnon coupling strength or by decreasing the magnon linewidth in the quantum regime~\cite{Tabuchi2014}. Together with the recently demonstrated bidirectional conversion between microwave and optical photons in yttrium iron garnet~\cite{Hisatomi2016}, this could pave the way to the transfer of quantum states between superconducting qubits and photons in optical fibers. Combining two very promising candidates for both stationary and flying qubits, such a breakthrough would be a important step toward the realization of a superconducting qubit-based quantum network.

\vspace{0.5cm}
\noindent\textbf{Acknowledgements}\newline

\noindent The authors acknowledge K. Usami, A. Blais, J. Bourassa, and M. Pioro-Ladri\`ere for useful discussions and P.-M. Billangeon for fabricating the transmon qubit. This work was partly supported by the Project for Developing Innovation System of MEXT, JSPS KAKENHI (Grant Number 26600071, 26220601), NICT, ERATO, JST, the Murata Science Foundation, Research Foundation for Opto-Science and Technology, Tateisi Science and Technology Foundation, JSPS Summer Program, and the Natural Sciences and Engineering Research Council of Canada (NSERC).\newline

\noindent\textbf{Authors contributions}\newline

\noindent  The measurements were carried out by D.L.-Q. and S.I., with input from all authors. D.L.-Q., Y.T., and Y.N. analysed the data and wrote the manuscript with input from all authors. Y.T., S.I., and Y.N. conceived the hybrid system with input from A.N., T.I., and R.Y.
\newline

\noindent\textbf{Additional information}\newline

\noindent Supplementary information is available in the online version of the paper. Reprints and permissions information is available online at www.nature.com/reprints. Correspondence and requests for materials should be addressed to D.L.-Q. and Y.N.\newline

\noindent\textbf{Competing financial interests}\newline

\noindent The authors declare no competing financial interests.

\end{document}


\title{Supplementary Information for ``Resolving magnon number states in quantum magnonics"}

\author{Dany Lachance-Quirion}
\affiliation{Institut quantique and D\'epartement de Physique, Universit\'e de Sherbrooke, J1K 2R1, Sherbrooke, Qu\'ebec,  Canada}
\affiliation{Research Center for Advanced Science and Technology (RCAST), The University of Tokyo, Meguro-ku, Tokyo 153-8904, Japan}
\author{Yutaka Tabuchi}
\author{Seiichiro Ishino}
\author{Atsushi Noguchi}
\author{Toyofumi Ishikawa}
\author{Rekishu Yamazaki}
\affiliation{Research Center for Advanced Science and Technology (RCAST), The University of Tokyo, Meguro-ku, Tokyo 153-8904, Japan}
\author{Yasunobu Nakamura}
\email{yasunobu@qc.rcast.u-tokyo.ac.jp}
\affiliation{Research Center for Advanced Science and Technology (RCAST), The University of Tokyo, Meguro-ku, Tokyo 153-8904, Japan}
\affiliation{Center for Emergent Matter Science (CEMS), RIKEN, Wako, Saitama 351-0198, Japan}

\maketitle

\tableofcontents
\newpage

\section{Experimental setup and hybrid system}

\begin{figure}[H]
\centering
\includegraphics[scale=0.95]{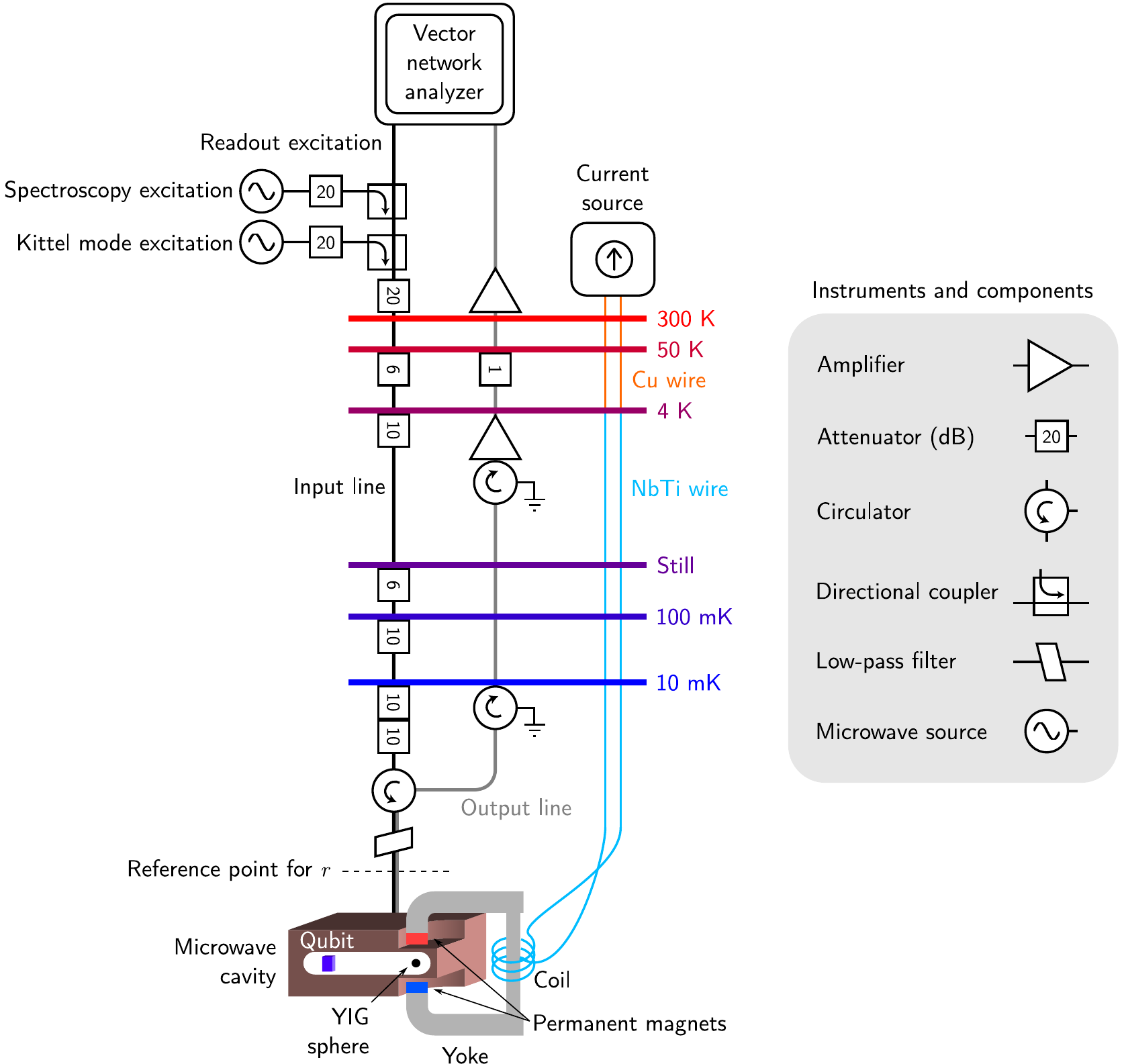}
\caption{\textbf{Experimental setup.} Spectroscopic measurements are performed with a vector network analyzer (Agilent E5071C). The microwave excitations are generated by microwave sources (Agilent E8247C), combined with the readout microwave excitation from the vector network analyzer in directional couplers (Krytar 120420), and introduced to the input port of the dilution refrigerator. The reflected signal from the cavity is amplified by amplifiers at 4~K (Caltech CRYO4-12) and at room temperature (MITEQ AFS4-08001200-09-10P-4). A current source (Yokogawa GS200) is used to supply the current $I$ to the coil through Cu wires (orange) and superconducting NbTi wires (pale blue). The attenuation of the input line (black) in the dilution refrigerator, of about $59$~dB at 10~GHz including losses in cables (phosphor-bronze coaxial cables; Coax Corp. SC-119/50-PBC- PBC) and connectors, is to prevent room-temperature thermal noise from reaching the hybrid system. A low-pass filter at 12~GHz (RLC F-30-12.4-R) is used to further decrease noise into the cavity. Noise from the room-temperature environment and the amplifiers in the output line (grey) are attenuated by more than $60$~dB by a circulator (Quinstar XTE0812KCS) and two isolators (Quinstar XTE0812KCS and XTE0812KC). A superconducting NbTi coaxial cable is used in the output line between the two isolators.
\label{fig:Figure_S1}}
\end{figure}

Figure~\ref{fig:Figure_S1} shows the instruments and components used in the experiment. Microwave powers $P_\mathrm{r}$, $P_\mathrm{s}$, and $P_\mathrm{mw}$ are calibrated using as the reference point the input of the cavity. At that reference point, the reflection coefficient $r$ is unity when $\left|\omega_\mathrm{r}-\omega_{10p}\right|\gg\kappa_{10p}$, where $\omega_\mathrm{r}$ is the readout frequency, and $\omega_{10p}$ and $\kappa_{10p}$ are the resonant frequency and the linewidth of the TE$_{10p}$ cavity mode, respectively, with $p=1,2,3\ldots$. Taking into account attenuation in cables outside and inside the dilution refrigerator, the total attenuation between the microwave sources and the input of the cavity are approximately $81$~dB, $122$~dB, and $121$~dB for the readout, spectroscopy, and Kittel mode microwave excitations, respectively.

The yoke, coil, cavity and YIG sphere of the hybrid system used in the paper are the same as in Ref.~\cite{Tabuchi2015a}, while the transmon qubit is a different one. The oxygen-free copper microwave cavity has dimensions of $24\times3\times53$~mm$^3$. A SMA connector connected to the cavity is used to measure the reflection coefficient $r$. A pair of disc-shape neodymium permanent magnets, with a diameter of 10~mm and a thickness of 1~mm each, are placed at the ends of a magnetic yoke made of pure iron. The magnets produce a static field $B\approx0.29$~T in the 4-mm gap between them. The magnetic field can be additionally tuned by a current $I$ in a $10^4$-turn superconducting coil. The field-to-current conversion ratio is approximately $1.7$~mT/mA. A YIG sphere glued to an aluminium-oxide rod along the $\langle110\rangle$ crystal axis is mounted in the cavity at the center of the gap between the magnets. The static field is applied in parallel with the $\langle100\rangle$ crystal axis. A transmon-type superconducting qubit, consisting of two large-area aluminium pads and a single Josephson junction (Al/Al$_2$O$_3$/Al), is fabricated on a silicon substrate and is mounted inside the cavity. The qubit and the YIG sphere are separated by 35~mm in the horizontal direction. A double-layer magnetic shield made of aluminium and pure iron covers half of the cavity to protect the qubit from the stray magnetic field of the magnet.

\section{Hamiltonian of the hybrid system}
\label{Section_parameters}

The Hamiltonian of the hybrid system is given by
\begin{align*}
\mathcal{\hat{H}}/\hbar&=\sum_{p=1}^\infty\omega_{10p}^\mathrm{bare}\hat a_p^\dagger\hat a_p+\left(\omega_\mathrm{q}^\mathrm{bare}-\alpha^\mathrm{bare}/2\right)\hat b^\dagger\hat b+\left(\alpha^\mathrm{bare}/2\right)\left(\hat b^\dagger\hat b\right)^2+\omega_\mathrm{m}^\mathrm{bare}\hat c^\dagger\hat c\numberthis\label{eq:Full_Hamiltonian}\\
&+\sum_{p=1}^\infty\left(g_{\mathrm{q},10p}\left(\hat a_p^\dagger\hat b+\hat a_p\hat b^\dagger\right)+g_{\mathrm{m},10p}\left(\hat a_p^\dagger\hat c+\hat a_p\hat c^\dagger\right)\right),
\end{align*}
where $\omega_{10p}^\mathrm{bare}$ is the bare frequency of the TE$_{10p}$ mode of the cavity, $\omega_\mathrm{q}^\mathrm{bare}\equiv\omega_{ge}$ and $\omega_{ef}$ are, respectively, the bare frequencies of the $|g\rangle\leftrightarrow|e\rangle$ and $|e\rangle\leftrightarrow|f\rangle$ transitions of the transmon qubit, $\alpha^\mathrm{bare}\equiv\omega_{ef}-\omega_{ge}$ is the bare anharmonicity of the transmon qubit, $\omega_\mathrm{m}^\mathrm{bare}$ is the bare magnon frequency, $g_{\mathrm{q},10p}$ is the coupling strength between the TE$_{10p}$ cavity mode and the transmon qubit, and $g_{\mathrm{m},10p}$ is the coupling strength between the TE$_{10p}$ cavity mode and the Kittel mode~\cite{Tabuchi2016}. In equation~\eqref{eq:Full_Hamiltonian}, $\hat a_p^\dagger~(\hat a_p)$, $\hat b^\dagger~(\hat b)$, and $\hat c^\dagger~(\hat c)$ are the creation (annihilation) operators of, respectively, a photon in the TE$_{10p}$ cavity mode, an excitation in the transmon qubit and a magnon in the Kittel mode. In the Hamiltonian of equation~\eqref{eq:Full_Hamiltonian}, the transmon qubit is considered as an anharmonic oscillator in order to take into account the effect of the $|e\rangle\leftrightarrow|f\rangle$ transition on the values of the calculated parameters, therefore capturing the straddling regime of the qubit-magnon system~\cite{Koch2007}.

The parameters of the hybrid system in equation~\eqref{eq:Full_Hamiltonian} are shown in Table~\ref{TableS1}. We calculate values of the qubit-magnon coupling strength $g_\mathrm{q-m}$, the qubit-TE$_{103}$ cavity mode dispersive shift $\chi_{\mathrm{q},103}$, the qubit-magnon dispersive shift $\chi_\mathrm{q-m}$, and the magnon Kerr coefficient $K_\mathrm{m}$ using these parameters and the above Hamiltonian by truncating the sum over the TE$_{10p}$ modes of the cavity to $p=4$. We consider the TE$_{10p}$ cavity mode number states $|n_ {10p}=\{0,1,2\}\rangle$, the transmon states $|i=\{g,e,f\}\rangle$, and the Kittel mode magnon number states $|n_\mathrm{m}=\{0,1,2\}\rangle$. More explicitly, we diagonalize the Hamiltonian and evaluate the parameters with
\begin{align*}
\chi_\mathrm{q,103}&=\frac{1}{2}\left(\omega_{103}^e-\omega_{103}^g\right),\\
\chi_\mathrm{q-m}&=\frac{1}{2}\left(\omega_\mathrm{m}^e-\omega_\mathrm{m}^g\right),\\
K_\mathrm{m}&=2\omega_{\mathrm{m},0\rightarrow1}^g-\omega_{\mathrm{m},0\rightarrow2}^g,
\end{align*}
where $\omega_{103}^{g(e)}$ is the frequency of the TE$_{103}$ cavity mode with the transmon in the ground (excited) state, $\omega_\mathrm{m}^{g(e)}$ is the frequency of the Kittel mode with the transmon in the ground (excited) state, and $\omega_{\mathrm{m},0\rightarrow n_\mathrm{m}}^g$ is the transition frequency of the Kittel mode between the magnon vacuum state and the $|n_\mathrm{m}\rangle$ magnon number state with the transmon in the ground state, with $\omega_{\mathrm{m}}^g\equiv\omega_{\mathrm{m},0\rightarrow1}^g$, such that for $K_\mathrm{m}=0$, $\omega_{\mathrm{m},0\rightarrow n_\mathrm{m}}^g=n_\mathrm{m}\omega_\mathrm{m}^g$. The qubit-magnon interaction strength $g_\mathrm{q-m}$ is simply calculated by half the splitting in the qubit-magnon hybridized energy levels. Figure~2b in the main text shows the calculated qubit-magnon dispersive shift $\chi_\mathrm{q-m}$ as a function of the dressed magnon frequency $\omega_\mathrm{m}^g$, while Fig.~\ref{fig:Figure_S6}a shows the magnon Kerr coefficient $K_\mathrm{m}$ as a function of the bare magnon frequency $\omega_\mathrm{m}^\mathrm{bare}$. Table~\ref{TableS4} summarizes the theoretical values of $g_\mathrm{q-m}$, $\chi_{\mathrm{q},103}$, $\chi_\mathrm{q-m}$, and $K_\mathrm{m}$.

As in the main text, the TE$_{102}$ and TE$_{103}$ cavity modes are, from now on, labeled the coupler and probe cavity modes, respectively. Therefore, indices `$102$' and `$103$' are replaced by indices `$\mathrm{c}$' and `$\mathrm{p}$', respectively.

\section{Cavity-magnon coupling}

\begin{figure}[H]
\centering
\includegraphics{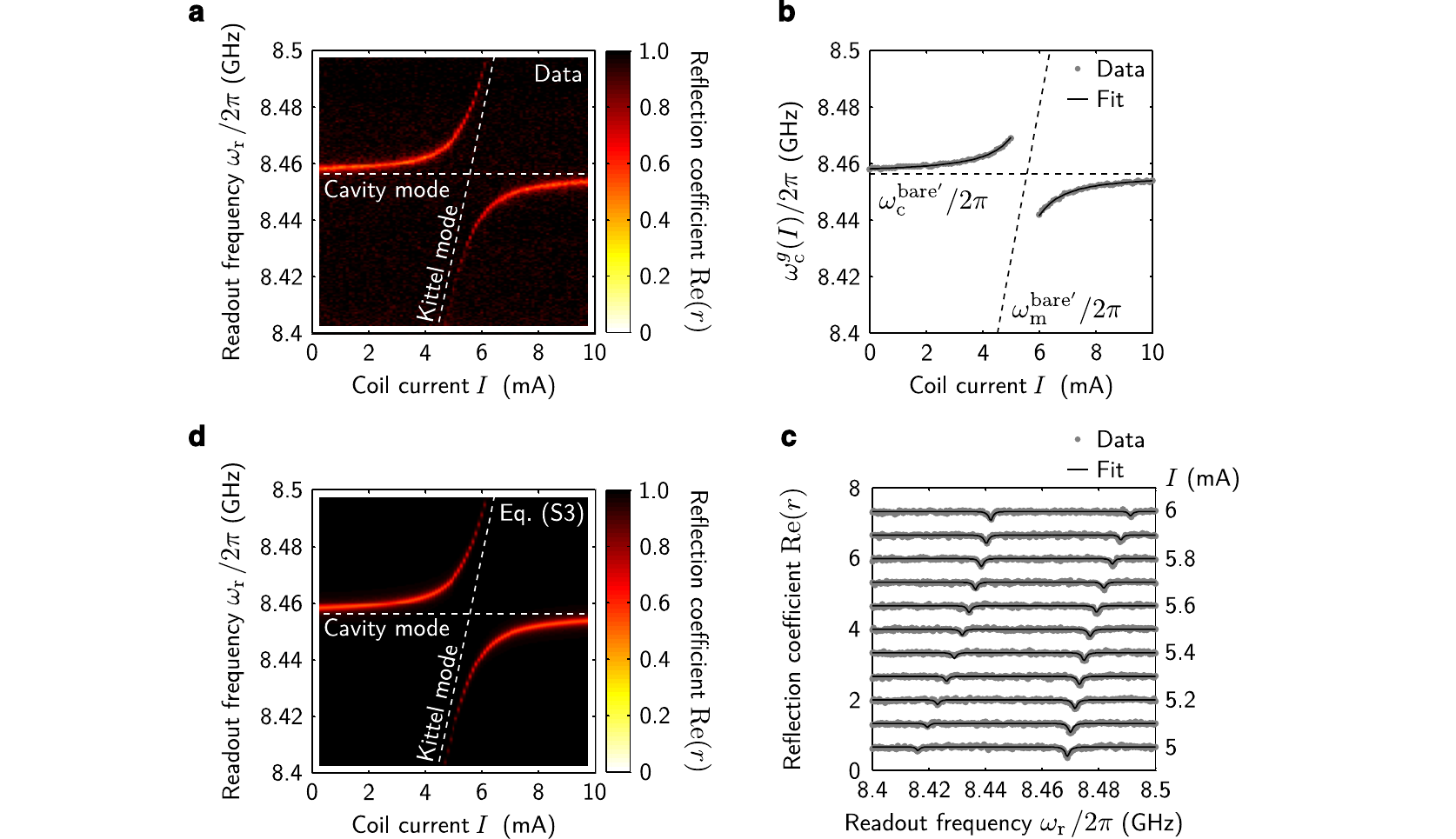}
\caption{\textbf{Cavity-magnon coupling.} \textbf{a}, Measurement of the reflection coefficient $\mathrm{Re}(r)$ of a readout microwave excitation at frequency $\omega_\mathrm{r}$ as a function of the coil current $I$. Both the spectroscopy and Kittel mode microwave excitations are turned off for this measurement. The avoided crossing in the coupler cavity mode spectrum indicates the coherent interaction between this cavity mode and the Kittel mode. \textbf{b}, Fit of the dressed frequency of the coupler cavity mode with the qubit in the ground state, $\omega_\mathrm{c}^g(I)$, to equation~\eqref{eq:Anticrossing_frequency}. \textbf{c}, Fit of the coupler cavity mode spectrum to equation~\eqref{eq:Anticrossing_spectrum} for different coil currents near the avoided crossing ($I=5$ to 6~mA). Individual spectra are offset vertically by $\mathrm{Re}(r)=1$ for clarity. \textbf{d}, Coupler cavity mode spectrum as a function of $I$ calculated using equation~\eqref{eq:Anticrossing_spectrum} with $g_\mathrm{m-c}/2\pi=22.5$~MHz and $\gamma_\mathrm{m}/2\pi=1.3$~MHz. For \textbf{a}, \textbf{b} and \textbf{d}, coupler and Kittel modes frequencies bare of their mutual interaction, $\omega_\mathrm{c}^{\mathrm{bare}^\prime}$ and $\omega_\mathrm{m}^{\mathrm{bare}^\prime}$ respectively, are shown as horizontal and diagonal dashed lines. 
\label{fig:Figure_S2}}
\end{figure}

Data on the avoided crossing between the $\mathrm{TE}_{102}$ cavity mode (coupler mode) and the Kittel mode is shown in Fig.~\ref{fig:Figure_S2}a. Figure~\ref{fig:Figure_S2}b shows the current-dependent dressed cavity frequency with the qubit in the ground state, $\omega_\mathrm{c}^g(I)$, extracted from data of Fig.~\ref{fig:Figure_S2}a. The dressed cavity frequency is fitted to
\begin{align}
\omega_\mathrm{c}^g(I)=p_1I+p_2-\mathrm{sgn}\left(I-I_0\right)\sqrt{\left(p_1I-p_3\right)^2+p_4^2},
\label{eq:Anticrossing_frequency}
\end{align}
with the fitting parameters $p_1$ to $p_4$ related to the physical quantities by
\begin{align*}
\omega_\mathrm{c}^{\mathrm{bare}^\prime}&=p_2+p_3,\\
\omega_\mathrm{m}^{\mathrm{bare}^\prime}(I)&=(p_2-p_3)+(2p_1)\times I,\\
\left|g_\mathrm{m-c}\right|&=p_4,\\
\omega_\mathrm{m}^{\mathrm{bare}^\prime}(I_0)&\equiv\omega_\mathrm{c}^{\mathrm{bare}^\prime}.
\end{align*}
In the above equations, $\omega_\mathrm{c}^{\mathrm{bare}^\prime}$ is the frequency of the coupler mode bare of its interaction with the Kittel mode, $\omega_\mathrm{m}^{\mathrm{bare}^\prime}$ is the frequency of the Kittel mode bare of its interaction with the coupler mode, $g_\mathrm{m-c}$ is the coupling strength between the coupler mode and the Kittel mode, and $I_0$ is the coil current for which $\omega_\mathrm{m}^{\mathrm{bare}^\prime}=\omega_\mathrm{c}^{\mathrm{bare}^\prime}$. This enables us to determine $\omega_\mathrm{c}^{\mathrm{bare}^\prime}/2\pi=8.456$~GHz.

We furthermore fit the cavity spectrum $\mathrm{Re}(r)$ to
\begin{align}
\mathrm{Re}(r)=\mathrm{Re}\left(\frac{\omega_\mathrm{r}-\omega_\mathrm{c}^{\mathrm{bare}^\prime}+\frac{i\left(\kappa_\mathrm{c}^\mathrm{int}-\kappa_\mathrm{c}^\mathrm{cpl}\right)}{2}-\frac{\left|g_\mathrm{m-c}\right|^2}{\omega_\mathrm{r}-\omega_\mathrm{m}^{\mathrm{bare}^\prime}(I)+i\gamma_\mathrm{m}/2}}{\omega_\mathrm{r}-\omega_\mathrm{c}^{\mathrm{bare}^\prime}+\frac{i\kappa_\mathrm{c}}{2}-\frac{\left|g_\mathrm{m-c}\right|^2}{\omega_\mathrm{r}-\omega_\mathrm{m}^{\mathrm{bare}^\prime}(I)+i\gamma_\mathrm{m}/2}}\right),
\label{eq:Anticrossing_spectrum}
\end{align}
where $\omega_\mathrm{r}$ is the readout frequency, $\kappa_\mathrm{c}^\mathrm{int}$ is the internal loss rate of the coupler mode, $\kappa_\mathrm{c}^\mathrm{cpl}$ is the coupling rate of the input/output port to the coupler mode, $\kappa_\mathrm{c}=\kappa_\mathrm{c}^\mathrm{int}+\kappa_\mathrm{c}^\mathrm{cpl}$ is the total linewidth of the coupler mode, and $\gamma_\mathrm{m}$ is the magnon linewidth~\footnote{The coupling rate of an unused port of the cavity is included in the internal loss rate of the cavity.}. Values of $\kappa_\mathrm{c}^\mathrm{int}$, $\kappa_\mathrm{c}^\mathrm{cpl}$, and  $\kappa_\mathrm{c}$, given in Table~\ref{TableS2}, are determined from a measurement of the coupler cavity mode spectrum far from the avoided crossing ($I=-10$~mA) while the value of $\omega_\mathrm{c}^{\mathrm{bare}^\prime}$ is fixed by the fit of $\omega_\mathrm{c}^g(I)$ to equation~\eqref{eq:Anticrossing_frequency}. The global fitting parameters are $\gamma_\mathrm{m}$ and $g_\mathrm{m-c}$ while $\omega_\mathrm{m}^{\mathrm{bare}^\prime}$ is fitted for each coil current $I$.

Figure~\ref{fig:Figure_S2}c shows spectra fitted to equation~\eqref{eq:Anticrossing_spectrum} for coil currents $I$ near the avoided crossing at $I_0\approx5.5$~mA. We find $g_\mathrm{m-c}/2\pi=22.5\pm0.1$~MHz and $\gamma_\mathrm{m}/2\pi=1.3\pm0.3$~MHz, with error bars corresponding to 95\% confidence intervals. The avoided crossing calculated with equation~\eqref{eq:Anticrossing_spectrum} and the parameters determined from the above fits is shown in Fig.~\ref{fig:Figure_S2}d.

\section{Qubit spectrum in the dispersive regime}

The effective Hamiltonian of a driven qubit-harmonic oscillator system in the dispersive regime is given by
\begin{align}
\mathcal{\hat{H}}/\hbar&=\frac{1}{2}\Delta_\mathrm{s}\hat\sigma_z+\left(\Delta_\mathrm{d}+\chi\right)\hat d^\dagger\hat d+\chi\hat\sigma_z\hat d^\dagger\hat d+\Omega_\mathrm{s}\left(\hat\sigma^-+\hat\sigma^+\right)+\Omega_\mathrm{d}\left(\hat d+\hat d^\dagger\right),
\label{eq:Dispersive_Hamiltonian}
\end{align} 
where $\Delta_\mathrm{s}=\omega_\mathrm{q}-\omega_\mathrm{s}$ is the spectroscopy detuning, $\omega_\mathrm{q}$ is the qubit frequency with the oscillator in the vacuum state, $\omega_\mathrm{s}$ is the spectroscopy excitation frequency, $\Delta_\mathrm{d}=\omega_\mathrm{o}^g-\omega_\mathrm{d}$ is the drive detuning, $\omega_\mathrm{o}^{g(e)}$ is the oscillator frequency with the qubit in the ground (excited state), $\omega_\mathrm{d}$ is the drive frequency, $\hat d^\dagger$ ($d$) is the creation (annihilation) operator of the oscillator, $\chi$ is the dispersive shift, $\Omega_\mathrm{s}$ is the spectroscopy excitation strength (Rabi frequency), $\Omega_\mathrm{d}$ is the oscillator excitation strength. In equation~\eqref{eq:Dispersive_Hamiltonian}, the qubit and the oscillator are in the frames rotating at the qubit frequency $\omega_\mathrm{q}$ and the oscillator frequency $\omega_\mathrm{o}^g$, respectively.

From the Hamiltonian of equation~\eqref{eq:Dispersive_Hamiltonian}, Gambetta \textit{et al.\hspace{-0.05cm}} obtained an analytical expression for the qubit spectrum $S(\omega_\mathrm{s})$~\cite{Gambetta2006}, given by
\begin{align}
S(\omega_\mathrm{s})=\sum_{n=0}^\infty \frac{1}{\pi}\frac{1}{n!}\mathrm{Re}\left(\frac{(-A)^{n}e^A}{\gamma_\mathrm{q}^{(n)}-i\left(\omega_\mathrm{s}-\omega_\mathrm{q}^{(n)}\right)}\right)\equiv\sum_{n=0}^\infty S_{n}(\omega_\mathrm{s}),
\label{eq:Spectrum_Gambetta}
\end{align}
with
\begin{align}
\omega_\mathrm{q}^{(n)}&=\omega_\mathrm{q}+B+n\left(2\chi+\Delta_\mathrm{d}\right)\label{eq:omega_q_n},\\
\gamma_\mathrm{q}^{(n)}&=\gamma_\mathrm{q}+\kappa\left(n+D^\mathrm{ss}\right)\label{eq:gamma_q_n},\\
A&=D^\mathrm{ss}\left(\frac{\kappa/2-i\left(2\chi+\Delta_\mathrm{d}\right)}{\kappa/2+i\left(2\chi+\Delta_\mathrm{d}\right)}\right),\\
B&=\chi(\overline{n}^g+\overline{n}^e-D^\mathrm{ss}),\\
D^\mathrm{ss}&=\frac{2(\overline{n}^g+\overline{n}^e)\chi^2}{(\kappa/2)^2+\chi^2+(\chi+\Delta_\mathrm{d})^2}\label{eq:D_s},\\
\overline n^g&=\frac{\Omega_\mathrm{d}^2}{(\kappa/2)^2+\Delta_\mathrm{d}^2}\label{eq:n_g},\\
\overline n^e&=\frac{\Omega_\mathrm{d}^2}{(\kappa/2)^2+(\Delta_\mathrm{d}+2\chi)^2}\label{eq:n_e}.
\end{align}
In the above equations, $\omega_\mathrm{q}^{(n)}$ and $\gamma_\mathrm{q}^{(n)}$ are respectively the frequency and the linewidth of the qubit peak corresponding to the number state $|n\rangle$ and the steady-state distinguishability $D^\mathrm{ss}$, and $\kappa$ is the linewidth of the harmonic oscillator. The steady-state distinguishability $D^\mathrm{ss}$ is defined as the separation between the steady-state coherent states $|\alpha_{g,e}^\mathrm{ss}\rangle$ created in the oscillator by the microwave excitation with the qubit in the ground state and the excited state, $D^\mathrm{ss}=\left|\alpha_e^\mathrm{ss}-\alpha_g^\mathrm{ss}\right|^2$. The last term in equation~\eqref{eq:gamma_q_n} shows that as the coherent states $|\alpha_{g,e}^\mathrm{ss}\rangle$ become more distinguishable, the qubit linewidth $\gamma_\mathrm{q}^{(n)}$ increases due to the measurement-induced dephasing~\cite{Gambetta2006}. The occupancy with the qubit in the ground (excited) state is given by $\overline{n}^{g(e)}=|\alpha_{g(e)}^\mathrm{ss}|^2=\langle\hat n\hat\Pi_\mathrm{q}^{g(e)}\rangle$, where $\hat\Pi_\mathrm{q}^{g(e)}=|g(e)\rangle\langle g(e)|$ is the projector of the qubit to its ground (excited) state.

For $\chi\gg\kappa$ and $\Delta_\mathrm{d}=0$, the steady-state distinguishability $D^\mathrm{ss}$ is simply given by $D^\mathrm{ss}=\overline n^g+\overline n^e$ and $A\rightarrow-D^\mathrm{ss}$. In that case, the component of the qubit spectrum with $n$ excitations, $S_{n}(\omega_\mathrm{s})$, has a Lorentzian lineshape. The qubit spectrum $S(\omega_\mathrm{s})$ is therefore well described by a sum of Lorentzian functions at frequencies $\omega_\mathrm{q}^{(n)}$ and of linewidths $\gamma_\mathrm{q}^{(n)}$ with a Poisson distributed spectral weight of mean given by $D^\mathrm{ss}$. However, for $\chi\sim\kappa$, $A$ becomes complex, leading to a non-Lorentzian lineshape for $S_{n}(\omega_\mathrm{s})$, with possibly negative values. However, the integral over $\omega_\mathrm{s}$ of the spectrum and its components is positive in all cases.

\section{Qubit spectroscopy - Magnon vacuum state}

\subsection{Measurement}

We perform spectroscopy of the qubit by probing the change $\Delta r$ in the reflection coefficient $r$ of a readout microwave excitation of fixed frequency $\omega_\mathrm{r}$ as a function of the spectroscopy frequency $\omega_\mathrm{s}$. For all qubit spectroscopy measurements presented here, $\omega_\mathrm{r}$ is fixed at the frequency of the dressed $\mathrm{TE}_{103}$ cavity mode (probe mode) with the qubit in the ground state at $\omega_\mathrm{p}^g/2\pi=10.44916$~GHz such that $\Delta_\mathrm{r}=\omega_\mathrm{p}^g-\omega_\mathrm{r}=0$. The readout excitation power $P_\mathrm{r}$ is fixed to $9.2$~aW, corresponding to an average number of photons in the probe mode much smaller than one. Indeed, the occupancy of the probe mode from the readout excitation is given by
\begin{align}
\overline{n}_\mathrm{p}^g=\frac{P_\mathrm{r}}{\hbar\omega_\mathrm{p}^g}\frac{\kappa_\mathrm{p}^\mathrm{cpl}}{(\kappa_\mathrm{p}/2)^2},
\label{eq:n_p}
\end{align}
where $\kappa_\mathrm{p}$ is the the total linewidth of the probe mode and $\kappa_\mathrm{p}^\mathrm{cpl}$ is the coupling rate of the cavity input-output port to the probe mode. With values of $\kappa_\mathrm{p}^\mathrm{cpl}$ and $\kappa_\mathrm{p}$ given in Tables~\ref{TableS2}, we obtain $\overline{n}_\mathrm{p}^g=0.078\pm0.004$ for $P_\mathrm{r}=9.2$~aW.

\subsection{Analytical model}

To take into account the finite occupancy of the probe mode for qubit spectra measured with the Kittel mode in the vacuum state ($P_\mathrm{mw}=0$), we use the analytical spectrum $S(\omega_\mathrm{s})$ of equation~\eqref{eq:Spectrum_Gambetta} by considering the probe cavity mode as the harmonic oscillator through the substitutions
\begin{align*}
n,\overline{n}^{g,e},\kappa&\rightarrow n_\mathrm{p},\overline{n}_\mathrm{p}^{g,e},\kappa_\mathrm{p},\\
\chi&\rightarrow\chi_\mathrm{q-p},\\
\Delta_\mathrm{d},\Omega_\mathrm{d}&\rightarrow\Delta_\mathrm{p},\Omega_\mathrm{p},\\
A,B,D^\mathrm{ss}&\rightarrow A_\mathrm{p},B_\mathrm{p},D_\mathrm{p}^\mathrm{ss}
\end{align*}
in Eqs.~\eqref{eq:omega_q_n} to \eqref{eq:n_e}. This leads to
\begin{align}
S(\omega_\mathrm{s})=\sum_{n_\mathrm{p}=0}^\infty \frac{1}{\pi}\frac{1}{n_\mathrm{p}!}\mathrm{Re}\left(\frac{(-A_\mathrm{p})^{n_\mathrm{p}}e^{A_\mathrm{p}}}{\gamma_\mathrm{q}^{(n_\mathrm{p})}-i\left(\omega_\mathrm{s}-\omega_\mathrm{q}^{(n_\mathrm{p})}\right)}\right)\equiv\sum_{n_\mathrm{p}=0}^\infty S_{n_\mathrm{p}}(\omega_\mathrm{s}).
\label{eq:Spectrum_Gambetta_p}
\end{align}
More explicitly, we fit the measured spectrum $\mathrm{Re}(\Delta r)$ to 
\begin{align}
\mathrm{Re}(\Delta r)=\mathcal{A}\sum_{n_\mathrm{p}=0}^{10} S_{n_\mathrm{p}}(\omega_\mathrm{s})+\mathrm{Re}(\Delta r)_\mathrm{off},
\label{eq:Fit_function_103}
\end{align}
where $\mathcal{A}$ is a conversion factor from $S(\omega_\mathrm{s})$ to $\mathrm{Re}(\Delta r)$, and $\mathrm{Re}(\Delta r)_\mathrm{off}$ is an offset of the spectrum from zero. The Fock basis of the probe mode is truncated to $n_\mathrm{p}=10$. The linewidth $\kappa_\mathrm{p}$ of the probe mode is fixed to the value determined from a fit of the spectrum of the probe mode (Table~\ref{TableS2}), and the readout detuning $\Delta_\mathrm{r}$ is zero. The fitting parameters are the qubit frequency with the probe mode in the vacuum state, $\omega_\mathrm{q}$, the qubit linewidth $\gamma_\mathrm{q}(P_\mathrm{s})$ broadened by the spectroscopy microwave excitation of power $P_\mathrm{s}$, the qubit-probe mode dispersive shift $\chi_\mathrm{q-p}$, the probe mode occupancy $\overline{n}_\mathrm{p}^g$ with the qubit in the ground state, the conversion factor $\mathcal{A}$, and the offset $\mathrm{Re}(\Delta r)_\mathrm{off}$.

\subsection{Fit}

\begin{figure}[H]
\centering
\includegraphics{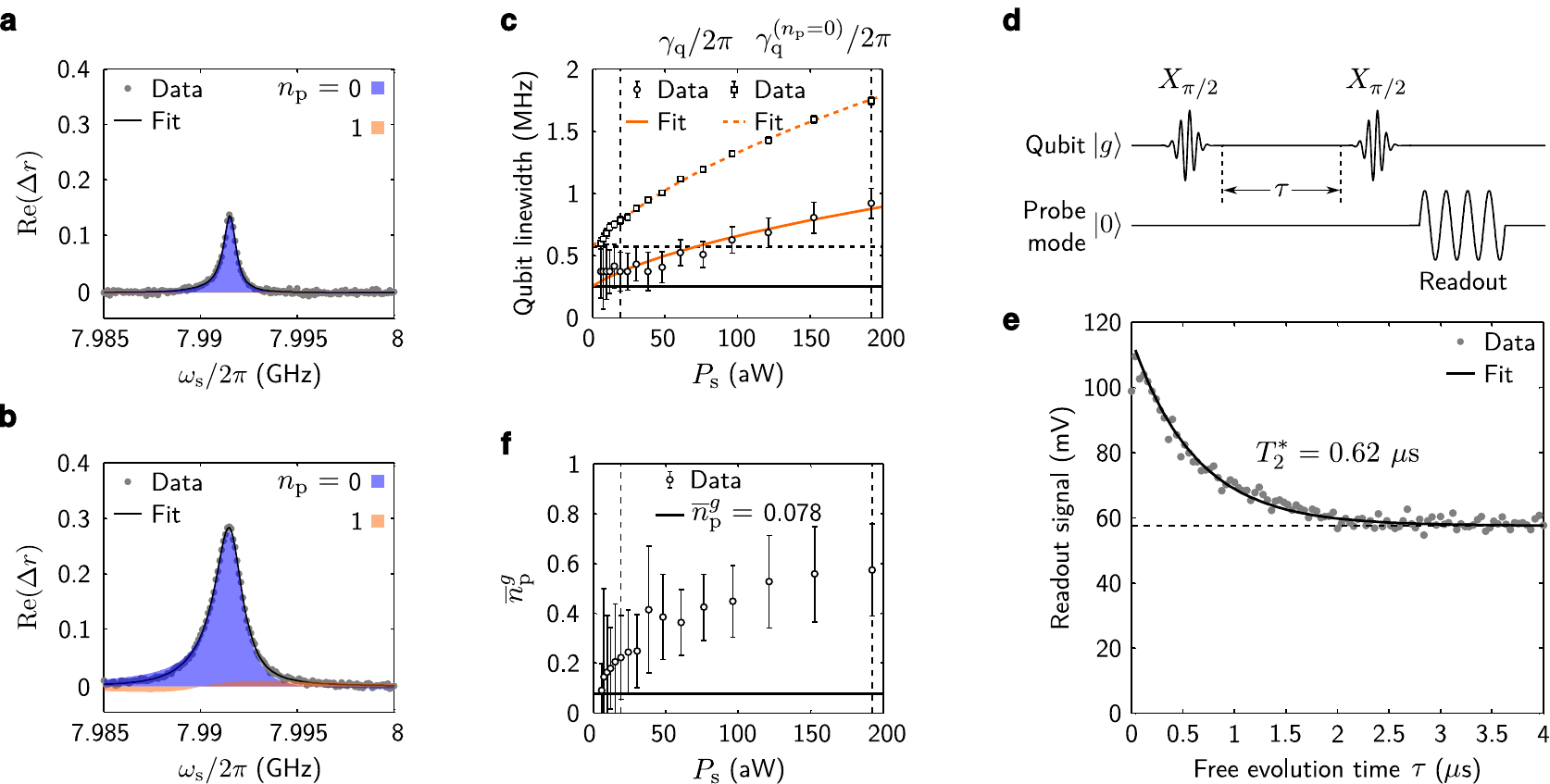}
\caption{\textbf{Power broadening of the qubit spectrum.} \textbf{a},~\textbf{b}, Qubit spectra for spectroscopy excitation powers $P_\mathrm{s}$ of \textbf{a}, 19~aW and \textbf{b}, 190~aW. Fits of data to equation~\eqref{eq:Fit_function_103} are shown as black lines. Blue and orange shaded areas respectively show the peaks corresponding to zero and one photons in the probe cavity mode. Note that the asymmetry in the qubit lineshape is very well reproduced by the fit. \textbf{c}, Power-broadened qubit linewidths $\gamma_\mathrm{q}$ (circles) and $\gamma_\mathrm{q}^{(n_\mathrm{p}=0)}$ (squares) in equation~\eqref{eq:gamma_q_n} as a function of $P_\mathrm{s}$. Solid and dashed orange lines show fits to equation~\eqref{eq_Power_broadening}, indicating linewidths $\gamma_\mathrm{q}(0)/2\pi=0.25$~MHz  (horizontal black solid line) and $\gamma_\mathrm{q}^{(n_\mathrm{p}=0)}(0)/2\pi=0.57$~MHz  (horizontal black solid line) for $P_\mathrm{s}\rightarrow0$. \textbf{d}, Pulse sequence used to measure the qubit dephasing time $T_2^*$ with Ramsey interferometry. An initial $\pi/2$ pulse prepares the qubit in a coherent superposition between the $|g\rangle$ and $|e\rangle$ states. After a time $\tau$ during which the qubit evolves freely, a second $\pi/2$ pulse is applied to the qubit. In the frame rotating at the qubit frequency, these two pulses would ideally result in the qubit in the excited state. In the presence of dephasing, the probability of finding the qubit in the excited state decays on a timescale given by the dephasing time $T_2^*$. Readout is performed by sending a strong microwave pulse resonant with the probe mode~\cite{Reed2010}. \textbf{e}, Readout signal as a function of the free evolution time $\tau$ between the two $\pi/2$ pulses. From the fit, we extract $T_2^*=0.62~\mu$s. The zero of the readout signal is defined as the signal with the qubit in the ground state, such that the readout signal is proportional to the probability $p_e$ of finding the qubit in the excited state after the pulse sequence shown in \textbf{d}. \textbf{f}, Occupancy of the probe cavity mode as a function of $P_\mathrm{s}$. Horizontal solid line shows the occupancy $\overline{n}_\mathrm{p}^g=0.078$ calculated with equation~\eqref{eq:n_p}. For \textbf{c} and \textbf{f}, vertical dashed lines show $P_\mathrm{s}=19$~aW and 190~aW. 
\label{fig:Figure_S3}}
\end{figure}

The measurements and the fits of the qubit spectra for spectroscopy excitation powers $P_\mathrm{s}$ of $19$~aW and $190$~aW are shown in Figs.~\ref{fig:Figure_S3}a and \ref{fig:Figure_S3}b. The dispersive shift $\chi_\mathrm{q-p}$ between the qubit and the probe mode is found to be $-0.8\pm0.2$~MHz, in excellent agreement with the expected value of $-0.73$~MHz~(Table~\ref{TableS4}). The power-broadened qubit linewidth $\gamma_\mathrm{q}(P_\mathrm{s})$, shown in Fig.~\ref{fig:Figure_S3}c, is fitted to
\begin{align}
\gamma_\mathrm{q}(P_\mathrm{s})=\sqrt{\eta P_\mathrm{s}+\gamma_\mathrm{q}(0)^2},
\label{eq_Power_broadening}
\end{align}
where $\eta\equiv(2\Omega_\mathrm{s})^2/P_\mathrm{s}$ relates $P_\mathrm{s}$ to the Rabi frequency $\Omega_\mathrm{s}$, and $\gamma_\mathrm{q}(0)$ is the intrinsic qubit linewidth~\cite{Schuster2005}. The Rabi frequency $\Omega_\mathrm{s}$ in equation~\eqref{eq:Dispersive_Hamiltonian} is estimated from the power-broadened qubit linewidth with
\begin{align}
\Omega_\mathrm{s}=\frac{1}{2}\sqrt{\gamma_\mathrm{q}(P_\mathrm{s})^2-\gamma_\mathrm{q}(0)^2}.
\label{eq:Rabi_frequency}
\end{align}

From the fit of $\gamma_\mathrm{q}(P_\mathrm{s})$ to equation~\eqref{eq_Power_broadening}, we find $\gamma_\mathrm{q}(0)/2\pi=0.25\substack{+0.07 \\ -0.10}$~MHz. To obtained this value, we restrict the minimal value of $\gamma_\mathrm{q}(P_\mathrm{s})$ in the fit such that the intrinsic linewidth respects the $T_1$ limit set by $\mathrm{min}\left[\gamma_\mathrm{q}(0)\right]=1/T_1$, with $T_1=0.63\pm0.07$~$\mu$s determined in a time-domain measurement. The linewidth $\gamma_\mathrm{q}^{(n_\mathrm{p}=0)}(0)=\gamma_\mathrm{q}(0)+\kappa_\mathrm{p}D_\mathrm{p}^\mathrm{ss}$ in equation~\eqref{eq:gamma_q_n}, corresponding to the linewidth of the peak with zero photons and broadened through measurement-induced dephasing from the photon occupancy of the probe mode, is $0.57\pm0.02$~MHz. The difference between $\gamma_\mathrm{q}(0)$ and $\gamma_\mathrm{q}^{(n_\mathrm{p}=0)}(0)$ can be explained by an occupancy in the probe mode of $0.20\substack{+0.20 \\ -0.09}$, significantly higher than the expected occupancy from the probe microwave excitation of $0.078\pm0.004$ previously estimated. This indicates a residual occupancy of $0.12\substack{+0.21 \\ -0.10}$, which should result in a linewidth of $0.44\substack{+0.54 \\ -0.27}$~MHz, broadened from the intrinsic linewidth even in the absence of both probe and spectroscopy microwave excitations. This linewidth compares well with the linewidth of $0.51\pm0.04$~MHz calculated with the dephasing time $T_2^*=0.62\pm0.04~\mu$s determined from Ramsey interferometry in a time-domain measurement~(Figs.~\ref{fig:Figure_S3}d and \ref{fig:Figure_S3}e). While all microwave excitations are turned off during the free evolution in the Ramsey interferometry measurement, the residual occupancy of the probe cavity mode creates measurement-induced dephasing of the qubit, increasing the linewidth from $0.25$ to $0.51$~MHz.

\section{Qubit spectroscopy - Magnon coherent state}

\subsection{Analytical model and fit}
\begin{figure}[H]
\centering
\includegraphics{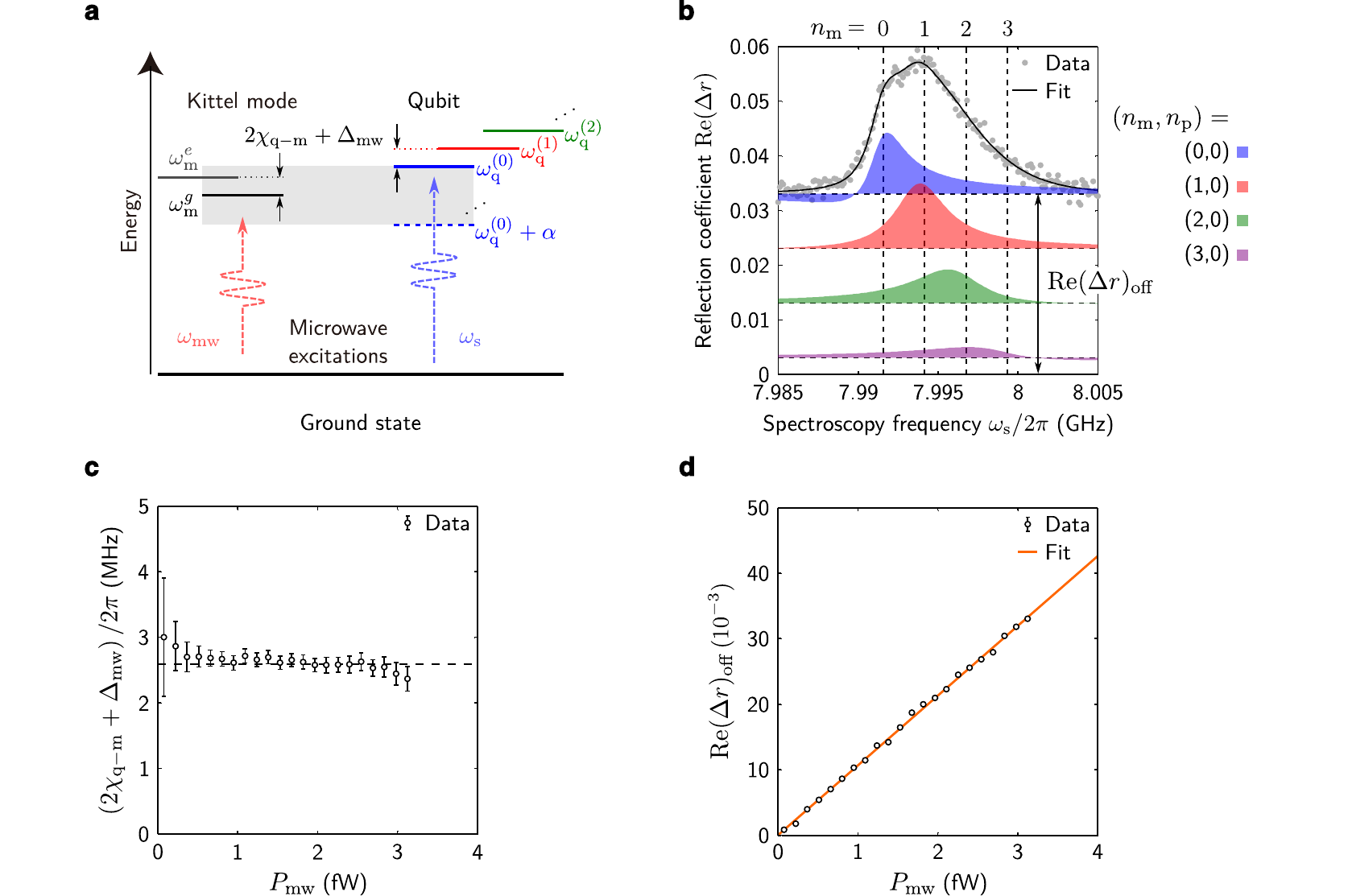}
\caption{\textbf{Dispersive qubit-magnon interaction.} \textbf{a},~Schematic energy diagram for the qubit and the Kittel mode in the dispersive regime, indicating the transitions $|g\rangle\leftrightarrow|e\rangle$ and $|e\rangle\leftrightarrow|f\rangle$ of the transmon at frequencies $\omega_\mathrm{q}^{(n_\mathrm{m})}$ and $\omega_\mathrm{q}^{(n_\mathrm{m})}+\alpha$ respectively, where $\alpha(<0)$ is the transmon anharmonicity and $|n_\mathrm{m}=\{0,1,2,\dots\}\rangle$ are the magnon number states. The straddling regime corresponds to $\omega_\mathrm{q}^{(0)}+\alpha<\omega_\mathrm{m}^g<\omega_\mathrm{q}^{(0)}$ (shaded area), where $\omega_\mathrm{m}^{g(e)}$ is the magnon frequency with the qubit in the ground (excited) state $|g(e)\rangle$. A microwave excitation at $\omega_\mathrm{mw}$ is used to drive the Kittel mode at a detuning $\Delta_\mathrm{mw}=\omega_\mathrm{m}^g-\omega_\mathrm{mw}$, leading to a splitting of $2\chi_\mathrm{q-m}+\Delta_\mathrm{mw}$ between qubit transitions corresponding to successive magnon number states. \textbf{b},~Qubit spectrum for $P_\mathrm{mw}=3.1$~fW and the corresponding fit (black line). Color-coded shaded areas show components of the spectrum corresponding to different magnon number states $|n_\mathrm{m}\rangle$ and probe mode in the vacuum state $|n_\mathrm{p}=0\rangle$. Components of the spectrum corresponding to one photon in the probe mode are not clearly visible and are therefore not shown. Components corresponding to $|n_\mathrm{m}=\{1,2,3\}\rangle$ are offset vertically by $-0.01$, $-0.02$ and $-0.03$, respectively, from the spectrum offset $\mathrm{Re}(\Delta r)_\mathrm{off}$. Negative values in the spectrum component corresponding to $n_\mathrm{m}=0$ are visible for $\omega_\mathrm{s}/2\pi\lesssim7.99$~GHz. \textbf{c},~Splitting between qubit transitions corresponding to successive magnon number states, $2\chi_\mathrm{q-m}+\Delta_\mathrm{mw}$, as a function of the Kittel mode excitation power $P_\mathrm{mw}$. At low excitation powers, it is difficult to determine from the splitting of the peaks the dispersive shift $\chi_\mathrm{q-m}$ and the excitation detuning $\Delta_\mathrm{mw}$ independently. We therefore omit data points for $P_\mathrm{mw}<0.9$~fW to estimate average values and standard deviations of $\chi_\mathrm{q-m}$ and $\Delta_\mathrm{mw}$. However, the splitting $2\chi_\mathrm{q-m}+\Delta_\mathrm{mw}$ is constant (within error bars) at $2.6\pm0.3$~MHz for all excitation powers. \textbf{d}, Offset $\mathrm{Re}(\Delta r)_\mathrm{off}$ as a function of the Kittel mode excitation power $P_\mathrm{mw}$. The orange solid line shows a linear fit of $\mathrm{Re}(\Delta r)_\mathrm{off}$. In \textbf{c} and \textbf{d}, error bars indicate 95\% confidence intervals. In \textbf{d}, error bars are smaller than the symbols.
\label{fig:Figure_S4}}
\end{figure}

Figure~\ref{fig:Figure_S4}a schematically represents the energy diagram of the qubit-magnon system in the dispersive regime. To fit the spectrum of the transmon qubit measured in this regime with a coherent excitation applied to the Kittel mode ($P_\mathrm{mw}>0$), we use the analytical spectrum $S(\omega_\mathrm{s})$ of equation~\eqref{eq:Spectrum_Gambetta} by considering the Kittel mode as the harmonic oscillator through the substitutions
\begin{align*}
n,\overline{n}^{g,e},\kappa&\rightarrow n_\mathrm{m},\overline{n}_\mathrm{m}^{g,e},\gamma_\mathrm{m},\\
\chi&\rightarrow\chi_\mathrm{q-m},\\
\Delta_\mathrm{d},\Omega_\mathrm{d}&\rightarrow\Delta_\mathrm{mw},\Omega_\mathrm{mw},\\
A,B,D^\mathrm{ss}&\rightarrow A_\mathrm{m},B_\mathrm{m},D_\mathrm{m}^\mathrm{ss}
\end{align*}
in Eqs.~\eqref{eq:omega_q_n} to \eqref{eq:n_e}. To take into account the ac Stark shift of the qubit frequency by the photons in the probe mode, we substitute
\begin{align}
\omega_\mathrm{q}\rightarrow\omega_\mathrm{q}^{(n_\mathrm{p}=0)}=\omega_\mathrm{q}+B_\mathrm{p},
\end{align}
where the ac Stark shifted qubit frequency with the Kittel mode in the vacuum state, $\omega_\mathrm{q}^{(n_\mathrm{p}=0)}/2\pi=7.99156$~GHz, is determined from the fit presented in Fig.~\ref{fig:Figure_S3}a. The qubit linewidth with the Kittel mode in the vacuum state is substituted to
\begin{align}
\gamma_\mathrm{q}\rightarrow\gamma_\mathrm{q}^{(n_\mathrm{p}=0)}=\gamma_\mathrm{q}+\kappa_\mathrm{p}D_\mathrm{p}^\mathrm{ss},
\end{align}
to take into account the increase in the linewidth by measurement-induced dephasing from photons in the probe mode, with $\gamma_\mathrm{q}^{(n_\mathrm{p}=0)}/2\pi=0.78$~MHz~(Fig.~\ref{fig:Figure_S3}c and Table~\ref{TableS3}). With theses substitutions, we fit the qubit spectrum to
\begin{align}
\mathrm{Re}(\Delta r)=\mathcal{A}\sum_{n_\mathrm{m}=0}^{10}S_{n_\mathrm{m}}(\omega_\mathrm{s})+\mathrm{Re}(\Delta r)_\mathrm{off},
\label{eq:Fit_function_m}
\end{align}
where, to take into account the asymmetry in the qubit lineshape from the qubit-probe mode dispersive interaction, we consider the one-photon peak of the probe mode with
\begin{align}
S_{n_\mathrm{m}}(\omega_\mathrm{s})\approx S_{n_\mathrm{m},n_\mathrm{p}=0}(\omega_\mathrm{s})+\mathcal{B}\times S_{n_\mathrm{m},n_\mathrm{p}=1}(\omega_\mathrm{s}),
\label{eq:Snm}
\end{align}
in equation~\eqref{eq:Fit_function_m}, where
\begin{align*}
\mathcal{B}\equiv\frac{p_{n_\mathrm{p}=1}}{p_{n_\mathrm{p}=0}}=\int\mathrm{d}\omega_\mathrm{s}\ \frac{S_{n_\mathrm{p}=1}(\omega_\mathrm{s})}{S_{n_\mathrm{p}=0}(\omega_\mathrm{s})}
\end{align*}
is the relative spectral weight between the one-photon and the zero-photon peaks. From measurements at $P_\mathrm{mw}=0$~(Fig.~\ref{fig:Figure_S3}a), we find $\mathcal{B}=0.03$, which is supposed to be constant in the following analysis. Figure~\ref{fig:Figure_S4}b shows an example of the qubit spectrum for a Kittel mode excitation power of 3.1~fW.

The parameters fixed in the fit of the qubit spectrum are the qubit frequency $\omega_\mathrm{q}^{(n_\mathrm{p}=0)}$, the power-broadened qubit linewidth $\gamma_\mathrm{q}^{(n_\mathrm{p}=0)}$, the magnon linewidth $\gamma_\mathrm{m}$, the relative spectral weight $\mathcal{B}$, the qubit-probe mode dispersive shift $\chi_\mathrm{q-p}$, and total linewidth $\kappa_\mathrm{p}$ of the probe mode. For each Kittel mode excitation power, the fitting parameters are the qubit-magnon dispersive shift $\chi_\mathrm{q-m}$ (Fig.~\ref{fig:Figure_S4}c), the Kittel mode excitation detuning $\Delta_\mathrm{mw}$ (Fig.~\ref{fig:Figure_S4}d), the magnon occupancy with the qubit in the ground state, $\overline n_\mathrm{m}^g$ (Fig.~3d in the main text), the conversion factor $\mathcal{A}$, and the offset $\mathrm{Re}(\Delta r)_\mathrm{off}$ (Fig.~\ref{fig:Figure_S4}f). Excitation-power-averaged values of $\Delta_\mathrm{mw}$ and $\chi_\mathrm{q-m}$, given in Tables~\ref{TableS3} and \ref{TableS4} respectively, are discussed in the main text.

From the value of $\Delta_\mathrm{mw}/2\pi=-0.38$~MHz, we can estimate the dressed magnon frequency $\omega_\mathrm{m}^g/2\pi=(\omega_\mathrm{mw}+\Delta_\mathrm{mw})/2\pi=7.94962$~GHz. The Lamb shift $(\omega_\mathrm{m}^\mathrm{bare}-\omega_\mathrm{m}^g)/2\pi=1.88$~MHz is calculated from diagonalization of the total Hamiltonian $\mathcal{\hat H}$ of equation~\eqref{eq:Full_Hamiltonian} using parameters of Table~\ref{TableS1}. This gives a bare magnon frequency $\omega_\mathrm{m}^\mathrm{bare}/2\pi=7.95150$~GHz, from which we calculate $\left|\omega_\mathrm{q}^\mathrm{bare}-\omega_\mathrm{m}^\mathrm{bare}\right|/2\pi=89$~MHz, much greater than the qubit-magnon coupling strength of 7.79~MHz.

The offset $\mathrm{Re}(\Delta r)_\mathrm{off}$ shows a linear scaling with $P_\mathrm{mw}$, with $\mathrm{Re}(\Delta r)_\mathrm{off}$ set to zero for $P_\mathrm{mw}=0$. The offset appears as a displacement of the reflection coefficient $\Delta r$ in phase space, indicating an ac Stark shift of the resonant frequency of the probe mode by the occupancy of the Kittel mode. In the detuning-dependent measurement shown in Fig.~2c of the main text, the Kittel mode spectrum is indeed visible through the dispersive interaction between the probe cavity mode and the Kittel mode~\cite{Haigh2015}.

\subsection{Kittel mode excitation}

In Fig.~3d of the main text, the magnon occupancy $\overline{n}_\mathrm{m}^g$ increases as a function of the Kittel mode excitation power at a rate of $0.342\pm0.008$ magnons per femtowatt. Theoretically, the magnon occupancy $\overline{n}_\mathrm{m}^g$ is given by
\begin{align}
\overline n_\mathrm{m}^g=\frac{\Omega_\mathrm{mw}^2}{(\gamma_\mathrm{m}/2)^2+\Delta_\mathrm{mw}^2},
\label{eq:n_m_g}
\end{align}
with the Kittel mode excitation strength $\Omega_\mathrm{mw}$ given from the input-output theory by
\begin{align}
\Omega_\mathrm{mw}=\sqrt{\frac{P_\mathrm{mw}}{\hbar\omega_\mathrm{mw}}}\sum_p \sqrt{\kappa_{10p}^\mathrm{cpl}}\left[\frac{g_{\mathrm{m},10p}}{\Delta_{\mathrm{m},10p}}+\frac{g_\mathrm{q-m}g_{\mathrm{q},10p}}{\Delta_\mathrm{q-m}\sqrt{\Delta_{\mathrm{m},10p}^2+\kappa_{10p}^2}}\right],
\label{eq:Omega_mw}
\end{align}
where $\Delta_{\mathrm{m},10p}=\omega_{10p}^\mathrm{bare}-\omega_\mathrm{m}^\mathrm{bare}$, $\Delta_\mathrm{q-m}=\omega_\mathrm{q}^\mathrm{bare}-\omega_\mathrm{m}^\mathrm{bare}$, and $\Delta_\mathrm{mw}=\omega_\mathrm{m}^g-\omega_\mathrm{mw}$. For a given $p$, the first term of equation~\eqref{eq:Omega_mw} describes the excitation of the Kittel mode through a virtual excitation in the $\mathrm{TE}_{10p}$ cavity mode, while the second term describes the excitation of the Kittel mode through a virtual excitation in the qubit excited by a virtual excitation in the $\mathrm{TE}_{10p}$ cavity mode. This leads to a slope of the magnon occupancy $\overline{n}_\mathrm{m}^g$ as a function of the excitation power $P_\mathrm{mw}$ given by
\begin{align}
\frac{\overline{n}_\mathrm{m}^g}{P_\mathrm{mw}}=\frac{1}{\hbar\omega_\mathrm{mw}}\frac{1}{(\gamma_\mathrm{m}/2)^2+\Delta_\mathrm{mw}^2}\left(\sum_p \sqrt{\kappa_{10p}^\mathrm{cpl}}\left[\frac{g_{\mathrm{m},10p}}{\Delta_{\mathrm{m},10p}}+\frac{g_\mathrm{q-m}g_{\mathrm{q},10p}}{\Delta_\mathrm{q-m}\sqrt{\Delta_{\mathrm{m},10p}^2+\kappa_{10p}^2}}\right]\right)^2,
\label{eq:slope}
\end{align}
Truncating to sum over cavity modes to $p=3$, we calculate $\overline{n}_\mathrm{m}^g/P_\mathrm{mw}=0.16\substack{+0.12 \\ -0.06}$~magnons per femtowatt with parameters given in Tables~\ref{TableS1} to \ref{TableS3}. Additionally, the linewidth and the coupling rate of the TE$_{101}$ cavity mode are $\kappa_{101}/2\pi=1.39$~MHz and $\kappa_{101}^\mathrm{cpl}/2\pi=0.13$~MHz, respectively. Error bars correspond to extremal values within the 95\% confidence interval of $\kappa_{102}^\mathrm{cpl}=\kappa_\mathrm{c}^\mathrm{cpl}$, $\kappa_{103}^\mathrm{cpl}=\kappa_\mathrm{p}^\mathrm{cpl}$, $\gamma_\mathrm{m}$, and $\Delta_\mathrm{mw}$. The discrepancy between the experimental and theoretical slopes of $\overline{n}_\mathrm{m}^g(P_\mathrm{s})$ of approximately a factor of two is most likely explained by an underestimation of $\Omega_\mathrm{mw}$ as not all excitation channels of the Kittel mode are taken into account.

\subsection{Probability distribution}
\begin{figure}[H]
\centering
\includegraphics{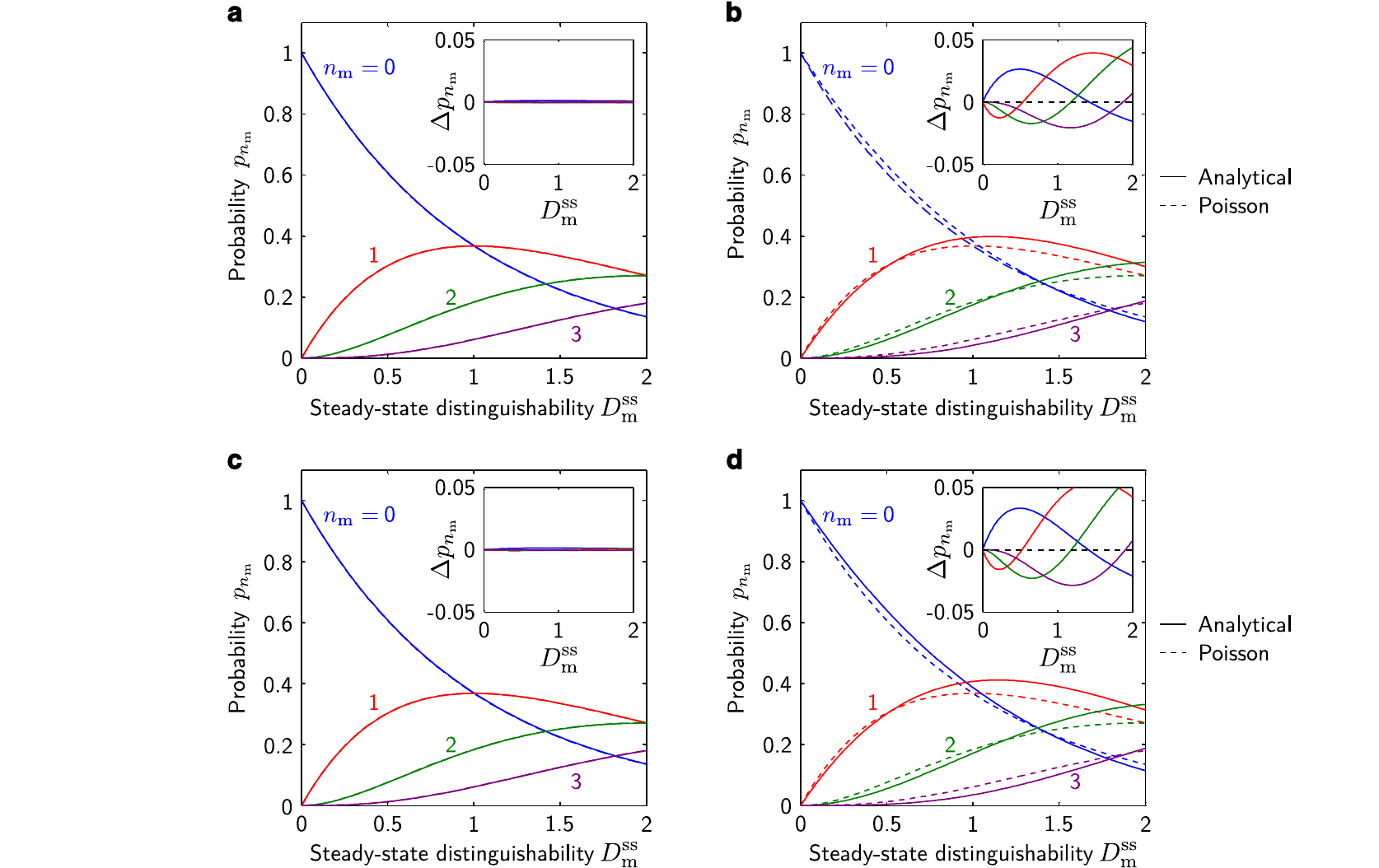}
\caption{\textbf{Probability distributions of magnon number states} calculated with equation~\eqref{eq:prob} (solid lines) and equation~\eqref{eq:prob_Poisson} (Poisson distribution, dashed lines) for \textbf{a}, $\gamma_\mathrm{m}/2\pi=0.1$~MHz and $\Delta_\mathrm{mw}=0$, \textbf{b}, $\gamma_\mathrm{m}/2\pi=1.3$~MHz and $\Delta_\mathrm{mw}=0$, \textbf{c}, $\gamma_\mathrm{m}/2\pi=0.1$~MHz and $\Delta_\mathrm{mw}/2\pi=-0.38$~MHz, and \textbf{d}, $\gamma_\mathrm{m}/2\pi=1.3$~MHz and $\Delta_\mathrm{mw}/2\pi=-0.38$~MHz. Insets show deviations $\Delta p_{n_\mathrm{m}}$ from Poisson distributions.
\label{fig:Figure_S5}}
\end{figure}

The probability $p_{n_\mathrm{m}}$ of the magnon number state $|n_\mathrm{m}\rangle$ is calculated as the relative spectral weight of each component $S_{n_\mathrm{m}}(\omega_\mathrm{s})$ of the fitted spectrum $S(\omega_\mathrm{s})$ with
\begin{align}
p_{n_\mathrm{m}}=\int\mathrm{d}\omega_\mathrm{s}\ \frac{S_{n_\mathrm{m}}(\omega_\mathrm{s})}{S(\omega_\mathrm{s})},
\label{eq:prob}
\end{align}
where
\begin{align*}
S(\omega_\mathrm{s})\approx\sum_{n_\mathrm{m}=0}^{10}S_{n_\mathrm{m}}(\omega_\mathrm{s}),
\end{align*}
with $S_{n_\mathrm{m}}(\omega_\mathrm{s})$ given by equation~\eqref{eq:Snm}. As discussed in Ref.~\cite{Gambetta2006}, the probability distributions of equation~\eqref{eq:prob} are to be compared with Poisson distributions of mean $D_\mathrm{m}^\mathrm{ss}$ given by
\begin{align}
p_{n_\mathrm{m}}=\frac{\left(D_\mathrm{m}^\mathrm{ss}\right)^{n_\mathrm{m}}e^{-D_\mathrm{m}^\mathrm{ss}}}{n_\mathrm{m}!}.
\label{eq:prob_Poisson}
\end{align}
Figure~\ref{fig:Figure_S5} shows the comparison between the Poisson distributions and the probability distributions calculated using equation~\eqref{eq:prob} with experimentally determined parameters, but with different values of the magnon linewidth $\gamma_\mathrm{m}$ and excitation detuning $\Delta_\mathrm{mw}$. For $\gamma_\mathrm{m}\ll\chi_\mathrm{q-m}$ ($\gamma_\mathrm{m}/2\pi=0.1$~MHz, Figs.~\ref{fig:Figure_S5}a and \ref{fig:Figure_S5}c), the probability distributions given by equation~\eqref{eq:prob} follow the Poisson distributions, even for a finite excitation detuning $\Delta_\mathrm{mw}$ of $-0.38$~MHz~\cite{Gambetta2006}. However, for $\gamma_\mathrm{m}\sim\chi_\mathrm{q-m}$ ($\gamma_\mathrm{m}/2\pi=1.3$~MHz, Figs.~\ref{fig:Figure_S5}b and \ref{fig:Figure_S5}d), systematic deviations from the Poisson distribution are observed. In that case, despite being in the strong dispersive regime with $\left|2\chi_\mathrm{q-m}\right|>\mathrm{max}\left[\gamma_\mathrm{q},\gamma_\mathrm{m}\right]$, the qubit does not perfectly probe the probability distribution of the Kittel mode. Error bars in Fig.~3e of the main text are calculated by finding the extremal values of $p_{n_\mathrm{m}}$ calculated within the 95\% confidence intervals of the fitting parameters $\overline n_\mathrm{m}^g$, $\chi_\mathrm{q-m}$, and $\Delta_\mathrm{mw}$.

\section{Magnon Kerr nonlinearity}

\begin{figure}[H]
\centering
\includegraphics{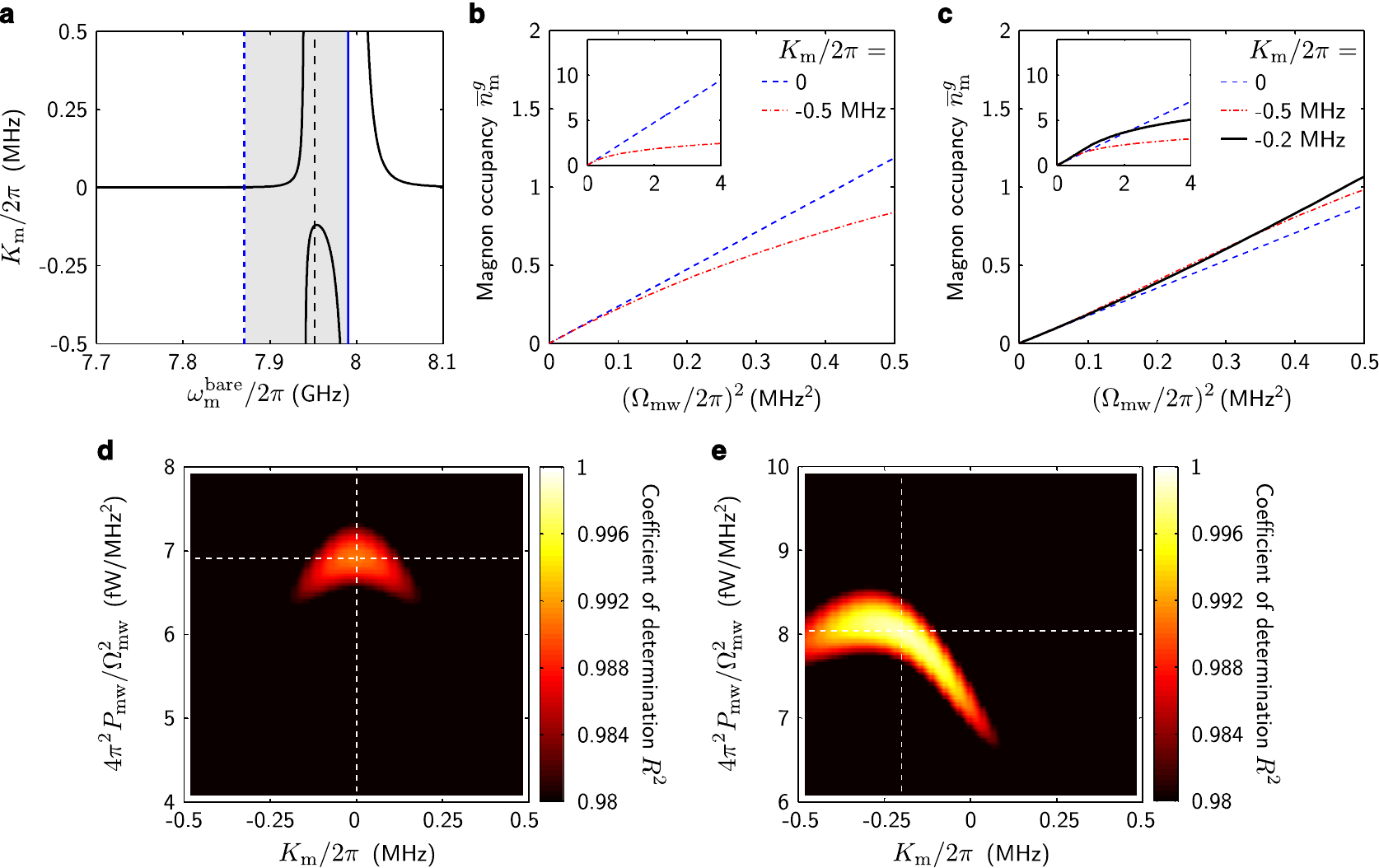}
\caption{\textbf{Magnon Kerr nonlinearity.} \textbf{a},~Calculation of the magnon Kerr coefficient $K_\mathrm{m}$ as a function of the bare magnon frequency $\omega_\mathrm{m}^\mathrm{bare}$ using the parameters of Table~\ref{TableS1}. Vertical black dashed line show $\omega_\mathrm{m}^\mathrm{bare}/2\pi=7.95150$~GHz, calculated from the experimentally determined $\omega_\mathrm{m}^g/2\pi=7.94962$~GHz and the calculated Lamb shift $(\omega_\mathrm{m}^\mathrm{bare}-\omega_\mathrm{m}^g)/2\pi=1.88$~MHz. The transmon qubit $|g\rangle\leftrightarrow|e\rangle$ and $|e\rangle\leftrightarrow|f\rangle$ dressed transition frequencies are shown as vertical solid and dashed blue lines, respectively. For a magnon frequency between these two frequencies, the qubit-magnon system is in the straddling regime (shaded area). \textbf{b,~c},~Numerical calculation of the magnon occupancy $\overline{n}_\mathrm{m}^g$ as a function of the Kittel mode excitation power, proportional to $\Omega_\mathrm{mw}^2$, using the Hamiltonian of equation~\eqref{eq:Magnon_Kerr} for different values of $K_\mathrm{m}$. The magnon linewidth $\gamma_\mathrm{m}$ is 1.3~MHz, and the Kittel mode excitation detuning $\Delta_\mathrm{mw}$ is zero in \textbf{b} and $-0.38$~MHz in \textbf{c}. Insets shows a larger range of Kittel mode excitation power. \textbf{d,~e},~ Coefficient of determination $R^2$ between $\overline{n}_\mathrm{m}^g(P_\mathrm{mw})$ (data, Fig.~3 in the main text) and $\overline{n}_\mathrm{m}^g(\Omega_\mathrm{mw})$ (simulations, this figure) as a function of $K_\mathrm{m}$ and the proportionality constant between $P_\mathrm{mw}$ and $\Omega_\mathrm{mw}^2$. The Kittel mode excitation detuning $\Delta_\mathrm{mw}$ is zero in \textbf{d} and $-0.38$~MHz in \textbf{e}. Vertical and horizontal dashed lines show best fit values of $K_\mathrm{m}$ and $4\pi P_\mathrm{mw}/\Omega_\mathrm{mw}^2$, respectively.
\label{fig:Figure_S6}}
\end{figure}

Using the Hamiltonian of the hybrid system of equation~\eqref{eq:Full_Hamiltonian}, we calculate the magnon Kerr coefficient $K_\mathrm{m}$ as a function of the bare magnon frequency $\omega_\mathrm{m}^\mathrm{bare}$ (Fig.~\ref{fig:Figure_S6}a). For $\omega_\mathrm{m}^g/2\pi=\left(\omega_\mathrm{mw}+\Delta_\mathrm{mw}\right)/2\pi=7.94962$~GHz at $I=-5.02$~mA, we estimate $K_\mathrm{m}/2\pi=-0.12$~MHz. As this coefficient is much smaller than the magnon linewidth of 1.3~MHz, it is not expected to significantly affect the dynamics of the Kittel mode.

However, to understand the effect of this nonzero Kerr coefficient on the behaviour of the magnon occupancy when increasing the Kittel mode excitation power, we consider the effective Hamiltonian of the driven qubit-magnon system in the dispersive regime given by
\begin{align*}
\mathcal{\hat{H}}_\mathrm{q-m}/\hbar&=\frac{1}{2}\Delta_\mathrm{s}\hat\sigma_z+\left(\Delta_\mathrm{mw}+\chi_\mathrm{q-m}+K_\mathrm{m}/2\right)\hat c^\dagger\hat c\numberthis\label{eq:Qubit-magnon_Hamiltonian}\\
&+\chi_\mathrm{q-m}\hat\sigma_z\hat c^\dagger\hat c-\left(K_\mathrm{m}/2\right)\left(\hat c^\dagger\hat c\right)^2\\
&+\Omega_\mathrm{s}\left(\hat\sigma^-+\hat\sigma^+\right)+\Omega_\mathrm{mw}\left(\hat c+\hat c^\dagger\right),
\end{align*} 
where $\Delta_\mathrm{s}=\omega_\mathrm{q}-\omega_\mathrm{s}$ is the spectroscopy detuning, $\Delta_\mathrm{mw}=\omega_\mathrm{m}^g-\omega_\mathrm{mw}$ is the Kittel mode excitation detuning, $K_\mathrm{m}$ is the coefficient of the magnon Kerr nonlinearity, $\Omega_\mathrm{s}$ is the spectroscopy excitation strength (Rabi frequency), and $\Omega_\mathrm{mw}$ is the Kittel mode excitation strength. In equation~\eqref{eq:Qubit-magnon_Hamiltonian}, the qubit and the Kittel mode are in the frames rotating at the dressed qubit frequency $\omega_\mathrm{q}$ and the magnon frequency $\omega_\mathrm{m}^g$, respectively.

By projecting $\mathcal{\hat H}_\mathrm{q-m}$ of equation~\eqref{eq:Qubit-magnon_Hamiltonian} into the $\hat\sigma_z\rightarrow-1$ subspace (qubit in the ground state), we obtain the Hamiltonian of a driven Kittel mode with a Kerr nonlinearity
\begin{align}
\mathcal{\hat{H}}_\mathrm{m}/\hbar=\left(\Delta_\mathrm{mw}+K_\mathrm{m}/2\right)\hat c^\dagger\hat c-\left(K_\mathrm{m}/2\right)\left(\hat c^\dagger\hat c\right)^2+\Omega_\mathrm{mw}\left(\hat c+\hat c^\dagger\right)\label{eq:Magnon_Kerr}.
\end{align}
Using Qutip~\cite{Johansson2012a,Johansson2013}, we numerically calculate the steady-state magnon occupancy $\overline{n}_\mathrm{m}^g$ with this Hamiltonian and under magnon relaxation at a rate $\gamma_\mathrm{m}/2\pi=1.3$~MHz, as a function of the Kittel mode excitation strength $\Omega_\mathrm{mw}$. The Kittel mode excitation detuning $\Delta_\mathrm{mw}$ is fixed to zero and $-0.38$~MHz in Figs.~\ref{fig:Figure_S6}b and \ref{fig:Figure_S6}c, respectively. At zero detuning ($\Delta_\mathrm{mw}=0$), a nonzero Kerr coefficient leads to negative curvature in $\overline{n}_\mathrm{m}^g(\Omega_\mathrm{mw}^2)$ as the excitation gets less efficient when the excitation strength increases, as the increasing magnon occupancy effectively change the excitation detuning for $|K_\mathrm{m}|>0$. However, for a finite detuning of $-0.38$~MHz, the curvature is positive at small $\overline{n}_\mathrm{m}^g$ for a finite range of Kerr coefficients $K_\mathrm{m}<0$. Indeed, in this case, the Kerr nonlinearity compensate the finite detuning, making the Kittel mode excitation more efficient as the magnon occupancy increases. For larger values of $\overline{n}_\mathrm{m}^g$, the curvature becomes negative, as shown in the inset of Fig.~\ref{fig:Figure_S6}c.

We now compare the numerically calculated $\overline{n}_\mathrm{m}^g(\Omega_\mathrm{mw}^2)$ to the data $\overline{n}_\mathrm{m}^g(P_\mathrm{mw})$ of Fig.~3d in the main text. Figures~\ref{fig:Figure_S6}d and \ref{fig:Figure_S6}e shows the coefficient of determination $R^2$ between $\overline{n}_\mathrm{m}^g(\Omega_\mathrm{mw}^2)$ and $\overline{n}_\mathrm{m}^g(P_\mathrm{mw})$ for different values of the Kerr coefficient $K_\mathrm{m}$ and the proportionally constant between $\Omega_\mathrm{mw}^2$ and $P_\mathrm{mw}$ Maximizing the coefficient of determination $R^2$, we determine a Kerr coefficient $K_\mathrm{m}$ of $-0.20\substack{+0.05 \\ -0.14}$~MHz for $\Delta_\mathrm{mw}/2\pi=-0.38$~MHz. Error bars on $K_\mathrm{m}$ are calculated by finding the extremal values found within the 95\% confidence interval of $\gamma_\mathrm{m}$ and $\Delta_\mathrm{mw}$. Even if the magnon Kerr coefficient is much smaller than the magnon linewidth, the value found from the fit is in relatively good agreement with the value of $-0.12$~MHz in Fig.~\ref{fig:Figure_S6}a. Finally, no thermal occupancy of the Kittel mode is found within our error bars of about 0.01 magnons, indicating an effective magnon temperature smaller than $\sim80$~mK.

\newpage

\newpage

\newpage

\newpage

\newpage

\begin{table*}[h!]
\begin{tabular}{l|c|c}
\hline 
Parameter & Symbol & Value (MHz)\\ \hline
\hhline{===}
$\mathrm{TE}_{101}$ cavity mode bare frequency & $\omega_{101}^\mathrm{bare}/2\pi$ & $6994.0$\\ \hline
$\mathrm{TE}_{102}$ cavity mode bare frequency & $\omega_{102}^\mathrm{bare}/2\pi=\omega_\mathrm{c}^\mathrm{bare}/2\pi$ & $8414.5$\\ \hline
$\mathrm{TE}_{103}$ cavity mode bare frequency & $\omega_{103}^\mathrm{bare}/2\pi=\omega_\mathrm{p}^\mathrm{bare}/2\pi$ & $10,441.5$\\ \hline
$\mathrm{TE}_{104}$ cavity mode bare frequency & $\omega_{104}^\mathrm{bare}/2\pi$ & $(12,800)$\\
\hhline{===}
Transmon bare $|g\rangle\leftrightarrow|e\rangle$ transition frequency & $\omega_\mathrm{q}^\mathrm{bare}/2\pi$ & $8040.6$\\ \hline
Transmon bare anharmonicity & $\alpha^\mathrm{bare}/2\pi$ & $-137.2$\\
\hhline{===}
$\mathrm{TE}_{101}$ cavity mode-qubit coupling rate & $g_{\mathrm{q},101}/2\pi$ & $73$\\ \hline
$\mathrm{TE}_{102}$ cavity mode-qubit coupling rate & $g_{\mathrm{q},102}/2\pi=g_\mathrm{q-c}/2\pi$ & $126.1$\\ \hline
$\mathrm{TE}_{103}$ cavity mode-qubit coupling rate & $g_{\mathrm{q},103}/2\pi=g_\mathrm{q-p}/2\pi$ & $135.4$\\ \hline
$\mathrm{TE}_{104}$ cavity mode-qubit coupling rate & $g_{\mathrm{q},104}/2\pi$ & $(116)$\\
\hhline{===}
$\mathrm{TE}_{101}$ cavity mode-Kittel mode coupling rate & $g_{\mathrm{m},101}/2\pi$ & $(-13.6)$\\ \hline
$\mathrm{TE}_{102}$ cavity mode-Kittel mode coupling rate & $g_{\mathrm{m},102}/2\pi=g_\mathrm{m-c}/2\pi$ & $22.5$\\ \hline
$\mathrm{TE}_{103}$ cavity mode-Kittel mode coupling rate & $g_{\mathrm{m},103}/2\pi=g_\mathrm{m-p}/2\pi$ & $(-20.3)$\\ \hline
$\mathrm{TE}_{104}$ cavity mode-Kittel mode coupling rate & $g_{\mathrm{m},104}/2\pi$ & $(14.0)$\\
 \hline
\end{tabular}
\caption{\textbf{Parameters of the hybrid system} used for the calculation of the qubit-magnon coupling strength $g_\mathrm{q-m}$, the qubit-probe mode dispersive shift $\chi_{\mathrm{q},103}=\chi_\mathrm{q-p}$, the qubit-magnon dispersive shift $\chi_\mathrm{q-m}$, and the magnon Kerr coefficient $K_\mathrm{m}$. Parameters in parentheses are numerically estimated based on electromagnetic field simulations.
\label{TableS1}}
\end{table*}

\begin{table*}[h!]
\begin{tabular}{l|c|c|c}
\hline 
Parameter & Symbol & Value (MHz) & Figure\\
\hhline{====}
Coupler cavity mode linewidth & $\kappa_\mathrm{c}/2\pi$ & $2.08\pm0.02$ & \\ \hline
Coupler cavity mode internal loss rate & $\kappa_\mathrm{c}^\mathrm{int}/2\pi$ & $1.58\pm0.02$ & -\\ \hline 
Coupling rate to the coupler cavity mode & $\kappa_\mathrm{c}^\mathrm{cpl}/2\pi$ & $0.51\pm0.02$ & -\\ \hline
\hhline{====}
Probe cavity mode linewidth & $\kappa_\mathrm{p}/2\pi$ & $3.72\pm0.03$ & -\\ \hline
Probe cavity mode internal loss rate & $\kappa_\mathrm{p}^\mathrm{int}/2\pi$ & $2.45\pm0.03$ & -\\ \hline 
Coupling rate to the probe cavity mode & $\kappa_\mathrm{p}^\mathrm{cpl}/2\pi$ & $1.27\pm0.03$ & -\\ \hline 
\hhline{====}
Intrinsic qubit linewidth & $\gamma_\mathrm{q}(0)/2\pi$ & $0.25\substack{+0.07 \\ -0.10}$ & \ref{fig:Figure_S3}c\\
\hhline{====}
Kittel mode linewidth & $\gamma_\mathrm{m}/2\pi$ & $1.3\pm0.3$ & \ref{fig:Figure_S2}\\
\hline
\end{tabular}
\caption{\textbf{Linewidths of the hybrid system.} Figures in Supplementary Information related to the parameters are indicated when available. Error bars indicate 95\% confidence intervals.
\label{TableS2}}
\end{table*}

\begin{table*}[h!]
\begin{tabular}{l|c|c|c|c|c}
\hline
Parameter & Symbol & Value in Fig.~1 & Value in Fig.~2 & Value in Fig.~3 & Figure\\
\hhline{======}
Readout excitation power & $P_\mathrm{r}$ & \multicolumn{3}{|c|}{9.2~aW} & -\\
\hline
Readout excitation frequency & $\omega_\mathrm{p}/2\pi$ & \multicolumn{3}{|c|}{10.44916~GHz} & -\\
\hline
Probe mode occupancy & $\overline{n}_\mathrm{p}^g$ & \multicolumn{2}{|c|}{$0.6\pm0.2$} & $0.22\pm0.17$ & \ref{fig:Figure_S3}f\\
\hhline{======}
Spectroscopy excitation power & $P_\mathrm{s}$ & \multicolumn{2}{|c|}{190~aW} & 19~aW & -\\
\hline
Broadened qubit linewidth & $\gamma_\mathrm{q}^{(n_\mathrm{p}=0)}(P_\mathrm{s})/2\pi$ & \multicolumn{2}{|c|}{$1.74\pm0.04$~MHz} & $0.78\pm0.03$~MHz & \ref{fig:Figure_S3}c\\
\hhline{======}
Kittel mode excitation power & $P_\mathrm{mw}$ & - & $7.9$~fW & $\left[0.079,3.1\right]$~fW & -\\
\hline
Kittel mode excitation detuning & $\Delta_\mathrm{mw}/2\pi$ & - & $\left[-10.38,4.62\right]$~MHz & $-0.38\pm0.08$~MHz & \ref{fig:Figure_S4}c\\
\hline
\end{tabular}
\caption{\textbf{Experimental parameters of the measurements presented in the figures of the main text.} Figures in Supplementary Information related to the parameters are indicated when available. Error bars indicate 95\% confidence intervals.
\label{TableS3}}
\end{table*}

\begin{table*}[h!]
\begin{tabular}{l|c|c|c|c|c}
\hline 
Parameter & Symbol & \multicolumn{2}{|c|}{Value (MHz)} & Error (\%) & Figure\\ \hline
 & & Experimental & Theoretical & &\\
\hhline{======}
Qubit-magnon coupling strength & $g_\mathrm{q-m}/2\pi$ & $7.79$ & $6.67$ & $+17$ & 1\\ \hline 
Qubit-probe mode dispersive shift & $\chi_\mathrm{q-p}/2\pi$ & $-0.8\pm0.2$ & $-0.73$ & $+7$ & \ref{fig:Figure_S3}\\ \hline 
Qubit-magnon dispersive shift & $\chi_\mathrm{q-m}/2\pi$ & $1.5\pm0.1$ & $1.27$ & $+18$ & 2, 3 and \ref{fig:Figure_S4}c\\ \hline 
Magnon Kerr coefficient & $K_\mathrm{m}/2\pi$ & $-0.20\substack{+0.09 \\ -0.13}$ & $-0.12$ & $+58$ & \ref{fig:Figure_S6}\\
\hline
\end{tabular}
\caption{\textbf{Comparison between experimental and theoretical values}, respectively determined from measurements and from diagonalization of the total Hamiltonian $\mathcal{\hat H}$ of equation~\eqref{eq:Full_Hamiltonian} using parameters of Table~\ref{TableS1}. The qubit-probe mode dispersive shift $\chi_\mathrm{q-p}$, the qubit-magnon dispersive shift $\chi_\mathrm{q-m}$, and the magnon Kerr coefficient $K_\mathrm{m}$ are evaluated at $\omega_\mathrm{m}^g/2\pi=\left(\omega_\mathrm{mw}+\Delta_\mathrm{mw}\right)/2\pi=7.94962$~GHz for $I=-5.02$~mA.
\label{TableS4}}
\end{table*}